\documentclass[12pt,preprint]{aastex}
\begin{document}

\title{A Candidate Young Massive Planet in Orbit around the Classical T
Tauri Star CI Tau$^1$}
%\title{A Search for A Young Massive Planet in Orbit around the T
%Tauri Star CI Tau$^1$}
%\title{Evidence for A Young Massive Planet in Orbit around the T Tauri Star CI Tau$^1$}

\author{Christopher M. Johns-Krull\altaffilmark{2,3}, Jacob N. McLane\altaffilmark{4,5}, L. Prato\altaffilmark{3,4},
Christopher J. Crockett\altaffilmark{3,6}, Daniel T. Jaffe\altaffilmark{7}, Patrick M. Hartigan\altaffilmark{2},
Charles A. Beichman\altaffilmark{8,9}, Naved I. Mahmud\altaffilmark{2}, Wei Chen\altaffilmark{2},
B. A. Skiff\altaffilmark{4}, P. Wilson Cauley\altaffilmark{2,10}, Joshua A. Jones\altaffilmark{2}, G. N. Mace\altaffilmark{7}}

\altaffiltext{1}{This paper includes data taken at The McDonald Observatory
of The University of Texas at Austin.}
\altaffiltext{2}{Department of Physics and Astronomy, Rice University,
MS-108, 6100 Main Street, Houston, TX 77005, USA}
\altaffiltext{3}{Visiting Astronomer at the Infrared Telescope Facility,
which is operated by the University of Hawaii under cooperative agreement
NCC 5-538 with the National Aeronautics and Space Administration, Office of
Space Science, Planetary Astronomy Program.}
\altaffiltext{4}{Lowell Observatory, 1400 West Mars Hill
  Rd. Flagstaff, AZ 86001, USA; jmclane@lowell.edu, lprato@lowell.edu}
\altaffiltext{5}{Department of Physics \& Astronomy, Northern Arizona University, S San Francisco Street, Flagstaff, AZ 86011, USA}
\altaffiltext{6}{Science News, 1719 N St NW, Washington, DC 20036, USA}
\altaffiltext{7}{Department of Astronomy, University of Texas, R. L. Moore
Hall, Austin, TX 78712, USA}
\altaffiltext{8}{Jet Propulsion Laboratory, California Institute of Technology, 4800 Oak Grove Drive, Pasadena, CA 91109, USA}
\altaffiltext{9}{NASA Exoplanet Science Institute (NExScI), California Institute of Technology, 770 S. Wilson Ave, Pasadena, CA 91125, USA}
\altaffiltext{10}{Department of Astronomy, Wesleyan University, 45 Wyllys Avenue, Middletown, CT 06459, USA}

\begin{abstract}

The $\sim$2 Myr old classical T Tauri star CI Tau shows periodic
variability in its radial velocity (RV) variations measured at infrared (IR)
and optical wavelengths.  We find that these observations are consistent with
a massive planet in a $\sim 9$-day period orbit.  These results are
based on 71 IR RV measurements of this system obtained
over 5 years, and on 26 optical RV measurements obtained over 9 years.
CI Tau was also observed photometrically in the optical
on 34 nights over $\sim$one month
in 2012.   The optical RV data alone are inadequate to identify 
an orbital period, likely the result of star spot and activity induced noise
for this relatively small dataset. 
The infrared RV measurements reveal significant 
periodicity at $\sim$9 days.  In addition, the full set of optical and 
IR RV
measurements taken together phase coherently and with equal 
amplitudes to the $\sim$9 day period.
Periodic radial velocity signals can
in principle be produced by cool spots, hot spots, and reflection of the
stellar spectrum off the inner disk, in addition to resulting from a planetary
companion.  We have considered each of these and find the planet hypothesis
most consistent with the data.  The radial velocity amplitude yields an $M \sin i$ 
of $\sim 8.1$ M$_{Jup}$; in conjunction with a 1.3 mm continuum emission 
measurement of the circumstellar disk inclination from the literature,
we find a planet mass of $\sim 11.3$ M$_{Jup}$, assuming 
alignment of the planetary orbit with the disk.

\end{abstract}

\keywords{planets and satellites: formation -- stars: individual (CI Tau) --
stars: low-mass -- stars: pre-main sequence -- starspots -- techniques: radial velocities}

\section{Introduction}

Exoplanetary systems are common and literally come in all sizes and 
configurations.  These span a parameter space that encompasses more apparently
stable arrangements than ever imagined for multi-planet systems, from the 
hyper-compact KOI-500 (Ragozzine et al. 2012) to the decades- and 
centuries-long orbits of the HR 8799 planets (Marois et al. 2010), in motion 
around a vast range of host stars.  Intriguingly, however, although 
exoplanetary systems are found around main sequence stars, post-main sequence
giants, brown dwarfs (Chauvin et al. 2004), intermediate age stars (Quinn et 
al. 2014), and even pulsars (Wolszczan \& Frail 1992), to date there are no confirmed radial velocity (RV)
detections of exoplanets caught in the act of formation around very young 
stars. 

There are good reasons for this.  Stellar systems presumably in the process 
of forming planets in circumstellar or circumbinary disks are typically 
located at relatively large distances, i.e. $>$120~pc.  The $\sim$10 Myr TW 
Hya region, at 50~pc, is much closer but contains only a few handfuls of 
young stars, many, but not all (Bergin et al. 2013), evolved beyond the 
planet-forming stage (Schneider et al. 2012).  Young moving groups near the 
Sun contain greater numbers of stars but are older yet, and, given their 
dispersion from their molecular cloud birthplaces, are not only more 
challenging to age-date, but are also mostly devoid of primordial 
planet-forming raw materials (Simon et al. 2012).  Tantalizingly, directly 
imaged planets in these moving groups are typically associated with processed
debris disks (e.g., Su et al. 2009; Apai et al. 2015), but the planetary 
bodies themselves have moved well past the formation stage.  Thus it is 
unknown precisely when and at what distances from the parent star planets 
form, how rapidly they migrate or are disrupted and/or ejected, and at what 
age planetary systems acquire stable configurations.

The obstacles to planet surveys around newly-formed stars in the closest 
regions, such as Ophiuchus and Taurus, are daunting, particularly in the case
of classical T Tauri stars (CTTSs), few Myr year old solar analogues with 
optically thick, actively accreting circumstellar disks.  Not only are these
stars correspondingly faint, but they are also among the most variable classes
of nearby object (Xiao et al. 2012; Stauffer et al. 2014).  From relatively
mild forms of variability, such as the changes originating from spots on the 
surface of a rapidly rotating star (e.g., Herbst et al. 2002), to clumpy accretion 
processes in the presence of strong magnetic fields (Graham 1992; Johns-Krull
et al. 1999), to extreme FU Ori behaviors and ideopathic outbursts/dimmings 
on the order of a visual magnitude or more detected on short time scales 
(e.g., Fischer et al. 2012; Hillenbrand et al. 2013).  In these environments, subtle observations of 
transits, direct imaging, and RV monitoring are fraught 
with complications.

Notwithstanding the challenges, impressive progress has been made in the 
search for young planets, largely through direct imaging studies \citep{neu05,
luh06, laf08, sch08, ire11, kra12, del13, bow13, kra14, sal15}.  The first putative 
imaged exoplanet, 2M1207b, was identified as a comoving companion to the 
substellar M8 dwarf 2M1207 in the TW Hya association (Chauvin et al. 2004, 
2005).  The LkCa 15 system, located in the younger Taurus region, was imaged
using non-redundant masking by Kraus \& Ireland (2012) who noted an unusual 
pattern of near- and mid-infrared (IR) emission in the inner hole of the 
LkCa 15 transition disk.  Several years of observations of LkCa 15 have 
revealed apparent orbital motion (Ireland \& Kraus 2014) and, more recently,
apparent accretion \citep{sal15} onto the candidate protoplanet.  Mass estimates 
for these objects come from comparing their estimated luminosity and 
temperature with theoretical evolutionary models.  However, such models 
are uncertain at these young ages and the observations required to determine
the luminosity and temperature with adequate precision are challenging and yield
considerable uncertainty in the final mass estimate for a given object. 
The companion to GQ Lup \citep{neu05} has mass estimates that
range from 1 M$_{JUP}$ to $\sim 40$ M$_{JUP}$ \citep{neu08}.  
In addition to direct imaging of potential planetary mass companions, some suggestive  
results have come from transit searches.  van Eyken et al. (2012) and 
Ciardi et al. (2015) observed transits potentially attributable 
to a planetary mass object in a $\sim$half day orbit around a $< 3$ Myr old 
T Tauri star in the Orion region, although these results have been called into question
by \citet{yu15} and \citet{how16}.

The relative lack of confirmed planetary
mass companions to very young stars may provide clues to the planet
formation process, or may simply be a testament to the difficulty in
finding young planets.  However, the transformational image of the HL Tau system taken 
in the recently commissioned long-baseline configuration with ALMA reveals 
numerous disk gaps highly suggestive of on-going planet formation at a very
young age (Partnership, ALMA, Brogan et al. 2015).  To fully understand the
planet formation and migration process, we will need to identify newly formed
planets around these young, difficult targets.  This particularly
includes looking for close in Jupiter mass and larger companions.  
Radial velocity surveys have revealed the existence of a brown dwarf 
desert (e.g., Marcy \& Butler 2000), an unexpected paucity of close brown
dwarf companions to solar-type stars. It is not yet known if this distribution
of secondaries is the result of the formation process itself or the result of 
evolution.  Armitage and Bonnell (2002) suggest that disks massive enough to 
form brown dwarf companions will be so massive that these companions
inevitably interact with the disk, migrate in, and merge with the central 
star.  In this case, close brown dwarf companions may be detected with higher 
frequency among young stars, particularly those still surrounded by massive 
disks.

In the last decade our team has undertaken an RV survey of $\sim$140 T Tauri 
stars in the Taurus region to look for signatures of RV variability indicative
of young, massive planets in short-period orbits.  Although we began our 
program at McDonald Observatory with high-resolution spectroscopy exclusively 
at optical wavelengths, we soon added high-resolution IR follow up 
spectroscopy at the NASA IRTF for candidate confirmation.  Visible light 
line bisector analysis, used to distinguish between the spot and companion 
origins of RV variability (e.g., Huerta et al. 2008; Prato et al. 2008), is
ineffective unless the $v \sin i$ of the star is significantly greater than
the spectrograph resolution element.  Thus for a star with $v \sin i$ values
$<$10 km~s$^{-1}$ and a resolution R$=$60,000 spectrograph, we obtain at most
4 resolution elements, insufficient for the line bisector analysis (Desort 
et al. 2007).  Because the photosphere-star spot contrast is reduced in the 
IR, the impact of star spots on IR RV observations is reduced by a factor of
at least four in the K band (Mahmud et al. 2011).  This serves as a key 
discriminant between the presence of spots or a planet, which the line 
bisector analysis failed to provide (Prato et al. 2008).  Hu\'elamo et al.
(2008) used this approach to show that the suspected planet around
TW Hya (Setiawan et al. 2008) was most likely a false signal produced by
spots on this rapidly rotating young star.

The current methodology for obtaining relatively high-precision IR RVs 
requires fitting observed or synthetic template spectra, representing the 
telluric spectrum, and a synthetic stellar photospheric spectrum to the observed 
spectrum of a candidate exoplanet host star in the 2.3~$\mu$m range.  This 
region of the K-band is rich in both deep CO $\Delta\nu=2$ lines for late 
type stars and in relatively deep lines of CH$_4$ in the Earth's atmosphere. 
Precisions of better than 100 m~s$^{-1}$ have been demonstrated by e.g.,  
Blake et al. (2010), Figueira et al. (2010), Bean et al. (2010),
Crockett et al. (2011), Bailey et al. (2012),  and
Davison et al. (2015) with this approach; relatively high RV precision is 
possible even for remarkably active T Tauri stars (Crockett et al. 2012).

For the RV standard star GJ 281, Crockett et al. (2012) used IR spectroscopy 
to identify an RMS scatter of 66 m~s$^{-1}$ in RV measurements taken with 
the CSHELL spectrograph on the NASA IRTF 3-meter over 48 epochs, and 30 
m~s$^{-1}$ with the NIRSPEC spectrograph on the Keck II telescope over 9 
epochs.  For CI Tau, a 2 Myr old, 0.80$\pm$0.02 M$_{\odot}$ (Guilloteau et 
al. 2014), classical T Tauri star with an actively accreting circumstellar 
disk, Crockett et al. found that the amplitude of the RV variations was
essentially the same in both the optical and K band, potentially suggestive 
of a planetary mass companion.  Using the 10 optical and 14 IR measurements
available at that time, Crockett et al. tentatively identified two periods
in the available signals; however, no significance or uncertainty in this periodicity
was identified.  Since the publication of Crockett et al.,
we have invested considerable effort in time domain observations of 
CI Tau to confirm this tentative result and determine the parameters
of the system.  The outcome of this extended investigation is presented here.
In Section 2 we describe our continued high-resolution spectroscopic 
observations at both optical and IR wavelengths, as well as our optical 
photometry.  Details of our light curve and RV analyses are provided in 
Section 3.  A discussion appears in Section 4 and we summarize our 
conclusions in Section 5.

\section{Observations and Data Reduction}

The observations and data reduction are described below.  All the reduced
data used in this paper are available for independent analysis at \hfill\break
http://torre.rice.edu/$\sim$cmj/CITau.

\subsection{IR}

\subsubsection{IRTF}
 
 Most of our IR RVs were taken with CSHELL (Tokunaga et
 al. 1990; Greene et al. 1993) at the 3~m NASA IRTF. CSHELL is a long-slit
 echelle spectrograph (1.08-5.5 $\mu$m) which utilizes a Circular
 Variable Filter (CVF), allowing isolation of a single order on a
 256$\times$256 InSb detector array. For our observations, the CVF was used to
 allow isolation of a 50 \AA\ spectral window centered at 2.298
 $\mu$m (order 25). This region is favorable for relatively precise spectroscopic analysis
 because of the presence of multiple photospheric absorption lines from the 2.293
 $\mu$m CO  $\nu = 2-0$ band head. Additionally, the presence of
 multiple strong telluric lines, predominantly CH$_{4}$ absorption
 features, allows us to use the Earth's atmosphere as a ``gas cell'' and
 imprints a relatively stable wavelength reference on our
 observations. The 0.5$''$ slit produced an average FWHM of 2.6 pixels
 (0.5 \AA, 6.5 km s$^{-1}$ , measured using comparison lamp spectra),
 corresponding to a spectral resolving power of R $\sim$46,000.

We obtained spectra of CI Tau on 34 nights between 2009 November and 2014 March (Table 1).
At the beginning of each night we took 20 flat fields and dark frames
along with six Ar, Kr, and Xe comparison lamp exposures to
create a wavelength zero point reference and dispersion solution.
All targets were observed using 10$''$
nodded pairs which enabled subtraction of sky emission, dark current,
and detector bias. Typical integration times were 600 s per
nod (Crockett et al. 2011).  Conditions permitting, on every night we obtained
spectra of the RV standard GJ 281 and telluric standards
with the identical set up except for shorter exposure times.

The data reduction strategy, implemented entirely in IDL, has been reported in our earlier publications (e.g., Crockett et al. 2012)
and follows that of Johns-Krull et al. (1999).  Median filtering of individual dark frames produced a master dark.
A nightly normalized flat field was created by averaging the flat field exposures, subtracting the master dark frame, and then
dividing by the mean of the dark-subtracted master flat.  Nodded pairs of target spectra were subtracted and the difference image
subsequently divided by the normalized flat field. We estimated read noise from the standard deviation of a
Gaussian fit to a histogram of the pixel values in the difference image ($\sim$30 e$^-$).  The curved spectral traces in the difference
image were fit with a second-order polynomial to identify the location of maximum and minimum flux along the dispersion direction.
For optimal spectral extraction, each nod pair was divided into four equally spaced bins of 64 columns along the dispersion direction.
Within each of these 64 column bins we constructed a 10$\times$ oversampled ``slit function'' (i.e. the distribution of flux in the
cross-dispersion direction). A rough estimate of the spectrum was created by summing the pixels in each column of the
difference image for each nod position. The limits included in this sum are from the midpoint between the two nod positions
on the detector to the edge of the area fully illuminated by the flat lamp, typically 6$-$70 pixels in each column for each nod position.
Each pixel in the bin was then sorted by its distance from the order center for the column the given pixel falls in. The flux in
each pixel was divided by the rough estimate of the spectrum for its appropriate column to normalize all the pixels
going into the slit function estimate. The flux in these offset ordered pixels was then median filtered with a seven-point moving box.
A flux estimate for each oversampled pixel was then determined by taking the median of all the pixels that fell in a given subpixel.
This then formed the oversampled master slit function. The multiple median filters generally remove the effects of cosmic rays
and uncorrected bad pixels on the determination of the slit function. We then fit this master slit function to a three Gaussian model: a
central Gaussian flanked by two satellite Gaussians. The amplitude, center, and width of each Gaussian were fit as free
parameters using the IDL implementation of the AMOEBA non-linear least-squares (NLLS) fitting algorithm (Nelder \& Mead 1965).
The resulting model was then normalized to unit area. This algorithm produces four model slit functions, one for each bin.
However, the actual slit function is a smoothly varying function of column number. Therefore, to smooth out the transitions from
bin to bin, 256 column slit functions were created by linearly interpolating between the four bin slit functions.

To determine the total flux in each column of the spectrum, we calculated the scale factor that best matches the model
slit function to the column data, per the recipe described in Horne (1986). In order to mask out spurious flux levels from
cosmic rays, an iterative sigma-clipping algorithm was implemented. This algorithm starts with an estimate of the total noise
from the measured read noise in the differenced image plus the Poisson noise from the target. We then subtracted our model fit
from the data in each detector column and masked those pixels for which the residual was 3-$\sigma$ greater than the estimated
noise.  One or two iterations were performed until the scale factor converged, thus providing an optimal value of the spectrum in
that column that is largely immune to hot pixels, cosmic rays, and other artifacts. This algorithm also provides an estimate of the
flux uncertainty at each location along the spectrum.

\subsubsection{Keck}

Three IR spectra were taken with NIRSPEC on the 10-meter Keck II telescope
in February and November of 2010 (Table 1).  NIRSPEC is a vacuum
cryogenic, high-resolution, cross-dispersed, near-IR spectrograph which
operates at the Nasmyth focus. For our
observations we used the N7 Filter (1.839-2.630 $\mu$m) with the
echelle and cross-disperser angles set 62.72 and 36.24 degrees,
respectively, providing imaging of orders 30 through 35 on the
1024$\times$1024 InSb detector. The $0.288''\times24''$ slit yielded a
median FWHM of 2.25 pixels (0.74 \AA, 9.6 km s$^{-1}$, measured from
lamp spectra), corresponding to a spectral resolving power R $\sim$ 31,000.
%To allow for optimal simultaneous coverage of both telluric
%and spectral lines, only order 33 (2.286-2.320 $\mu$m) was utilized for our analysis.

Multiple flat, dark, and comparison lamp frames of Ne, Ar, Xe, and Kr were taken
on every night of observation.  The comparison lamp lines provided an initial
wavelength zero-point and dispersion solution. All targets were observed using a 10$''$
ABBA nod pattern, allowing for the subtraction of sky emission.
Target integration times were typically on the order of 30~s with 2$-$3 coadds. 
On all three nights on which we obtained CI Tau spectra we observed the RV standard GJ 281 and telluric standards;
integration times were 10$-$20s with 2$-$6 coadds.

The same data reduction procedure was applied as for the CSHELL observations (\S2.2.1) except that a fourth-order polynomial 
was used to trace the location of the spectra on the detector instead of a second-order polynomial.
Reductions  were limited to spectral order 33 (2.286$-$2.320 $\mu$m) because it contains the requisite
stellar CO and telluric CH$_4$ lines and encompasses the CSHELL bandpass
which allows for more direct comparison of the two instruments.

\subsubsection{KPNO}

Observations of CI Tau and GJ 281 were obtained with the Phoenix IR
echelle spectrometer (Hinkle et al. 1998) during four separate observing
runs in 2013 and 2014.  Data were obtained at the KPNO 4~m Mayall telescope
from 27 February 2013 through 3 March 2013.  All other KPNO data were obtained
at the 2.1~m telescope between 9 November 2013 and 10 January 2014 (Table 1).  At both
telescopes the four-pixel slit was used, corresponding to 
$0.^{\prime\prime}7 \times 28^{\prime\prime}$ at the Mayall 4~m and 
$1.^{\prime\prime}4 \times 56^{\prime\prime}$ at the KPNO 2.1~m.
The grating was configured to provide wavelength coverage
from 2.2943 -- 2.3040 $\mu$m, and the K4308-order blocking filter
was used to eliminate light from any overlapping orders.
This setup yielded a spectral resolution of $R \sim 50,000$ at both
telescopes.  Observations of each star were taken in pairs with a nod of
10$^{\prime\prime}$ along the slit for the Mayall 4~m telescope and a
nod of 20$^{\prime\prime}$ along the slit at the KPNO 2.1~m telescope.
Signal-to-noise ratios (SNR) varied significantly for the Phoenix observations
depending on the combination of object brightness, telescope used, and
observing conditions for the observation.  At the Mayall 4~m, typical 
observations on CI Tau consisted of four 600 s exposures, while for GJ 281 two
600 s exposures were most often used.  At the KPNO 2.1~m, eight 900 s observations
were typically made of CI Tau, while four 900 s exposures of GJ 281 were
sufficient.  A total of 19 observations of CI Tau and 9 of GJ 281were taken.
In addition to stellar spectra, flat field and dark exposures were also obtained,
as well as exposures of a ThArNe lamp in order to provide wavelength 
calibrations.  All of the data were reduced using custom IDL routines. These are essentially
the same routines used to reduce the CSHELL data, with small modifications
made to account for the differences in detector size and data formats of
each instrument.  

%  CHRIS -- I REMOVED THE FOLLOWING BC IT IS DESCRIBED IN DETAIL UNDER CSHELL BUT NOT UNDER NIRSPEC;
%  I ALSO REMOVED THE  HINKLE + 2001 REFERENCE FROM THE REF LIST.

%Briefly, each pair of images was differenced and 
%flat-fielded. The differenced spectra were then optimally extracted and 
%coadded after being interpolated onto the same wavelength scale.   The spectra
%from multiple pairs of exposures were also coadded to get one final spectrum
%per night.  Any remaining bad pixels (e.g. hot columns, cosmic rays that were
%not removed by the optimal extraction) were manually averaged between adjacent
%pixels.  Third-order polynomial wavelength solutions were
%obtained by fitting the observed ThArNe lamp exposures using
%line identifications from Hinkle et al. (2001).

\subsubsection{McDonald Observatory}

We obtained 14 observations of CI Tau on the Harlan J. Smith 2.7-meter telescope with the Immersion GRating INfrared Spectrograph (IGRINS) in 
2014 November and December and 2015 January (Table 1).
IGRINS implements silicon immersion gratings, a fixed optical path (no cryogenic mechanisms), and volume-phase holographic gratings to 
simultaneously cover the H and K bands (1.48-2.48$\mu$m) with a resolving power R$\sim$45,000. 
The echellogram for each band is projected onto a pair of 2048$\times$2048 pixel Teledyne H2RG HgCdTe detectors.
IGRINS straightforward design and high throughput make observations on the 2.7~m Harlan J. Smith Telescope at McDonald
Observatory comparable to spectrographs at  8~m facilities, but with 5 times to $\>$100 times the spectral grasp. 
Additional discussion on the design and capabilities of IGRINS appears in \citet{park2014}.

Observations with IGRINS employed standard near-IR techniques. 
CI Tau was nodded between two positions, separated by 7$\farcs$0, along the 1x15 arcsecond slit.
The IGRINS pipeline package PLP\footnote{Currently available at: https://github.com/igrins/plp} was developed 
by Dr. Jae-Joon Lee at Korea Astronomy and Space Science Institute and Professor Soojong Pak's team at Kyung Hee University.
The pipeline subtracts AB pairs to remove OH emission lines and then optimally extracts the sources based on the methods of \citet{horne1986}.
The wavelength solution is determined first from ThAr lamp spectra taken at the start of each night, and then
improved by fitting the OH lines in the two dimensional target spectra.
The flux, wavelength, signal-to-noise and variance of every extracted order is output as a FITS file for the H and K bands separately.
For our determination of CI Tau RVs we employed the CO bandhead lines redward of $\sim$2.295 $\mu$m.

\subsection{Optical}

\subsubsection{Echelle Spectroscopy}

CI Tau was included among the earliest subset of targets observed in our McDonald Observatory RV survey
of $\sim 140$ stars in the Taurus region to look for evidence of young, very
low-mass companions to newly formed stars.  The observational setup has been 
described in previous papers from this work (Huerta et al. 2008;
Prato et al. 2008; Mahmud et al. 2011; Crockett et al. 2012); we provide a brief summary here.
Spectra were obtained with the Robert G. Tull Coud\'e 
Spectrograph (Tull et al. 1995) at the McDonald Observatory 2.7~m Harlan J. Smith telescope.
A total of 29 spectra were obtained 
between 2004 December  28 and 2013 November 15 (Table 2).  A 
1.$^{\prime\prime}$2 slit yielded a spectral resolving power
of $R \equiv \lambda/\Delta\lambda \sim 60,000$ for spectra covering the
wavelength range 3,900 -- 10,000 \AA\ with small wavelength gaps between the
redder orders starting at $\sim 5,600$ \AA. Integration times varied
from 1200 s to 3000 s, depending on conditions,  but were usually
2400 s.  The average seeing was $\sim 2^{\prime\prime}$. We took
ThAr lamp exposures before and after each spectrum for wavelength
calibration; typical rms values for the dispersion solution
precision were $\sim 4$ m s$^{-1}$. 

The optical echelle spectra were reduced using a suite of IDL routines that
have been described in various references (e.g., Valenti 1994; Hinkle et al.
2000).  These routines form the basis of the REDUCE IDL echelle reduction 
package (Piskunov \& Valenti 2002). The raw spectra were bias-subtracted using
the overscan region and flat-fielded using the spectrum of an
internal continuum lamp. Optimal extraction to remove cosmic
rays and improve signal was used for all the spectra. The wavelength solution
was determined by fitting a two-dimensional polynomial to $n\lambda$ as 
function of pixel and order number, $n$, for approximately 1800 
extracted thorium lines observed from the internal lamp assembly.
The final wavelength solution used for each observation was
the average of solutions from ThAr lamp exposures taken before and after 
each stellar exposure.

\subsubsection{Photometry}

The photometry was obtained with the Lowell 0.7~m f/8 reflector
in robotic mode.  It has a permanently mounted CCD camera that provides
a $15'\times15'$ field at an image scale of 0$\farcs$9 pixel$^{-1}$.  
The CI Tau field was
targeted on fourteen nights between 2012 November 7 and December 11 UT.
We obtained two 3-minute exposures in the V filter at each visit,
with several visits each night yielding 189 data points (Table 3).
The images were reduced via ordinary aperture photometry with the 
commercial photometry package {\it Canopus}\, (version 10.4.0.6).
The four comparison stars were found to be constant over the
observation interval.  V magnitudes for these standard stars were adopted from
ASAS-3 (Pojmanski 1997) and APASS (Henden \& Munari 2014) to adjust
the data approximately to standard V magnitudes.  Because of the 
emission-line nature of the CI Tau spectrum, there will inevitably
be a small zero-point shift dependent on the color of the comparison stars
and the passband of the filter $+$ CCD system.  Our mean magnitude
near V$=$13.0 is nevertheless similar to the longer-term ASAS-3
value (V$=$13.04), the TASS MkIV series (V$=$13.1; Droege et al. 2006), 
and APASS with sparser observations (V$=$12.94).

\section{Analysis}

\subsection{IR RVs}

All near-IR K-band observations were processed in essentially the same way
using procedures described by Crockett et al. (2011). 
The RVs were determined using a spectral modeling technique in which
template spectra for the stellar spectrum and the telluric spectrum are
combined to model each of the observed spectra of CI Tau. We interpolated
over a grid of NextGen models (Hauschildt et al. 1999) to
produce a synthetic stellar atmosphere tailored to the $T_{eff}$, log$g$, and metallicity 
assumed for CI Tau. We then used SYNTHMAG (Piskunov 1999) to create a 
template stellar spectrum with this model atmosphere using an atomic line
list from Kupka et al. (2000) and a CO line list from Goorvitch (1994).
For the telluric spectral template, we used the NOAO telluric absorption 
spectrum of Livingston \& Wallace (1991).
The telluric absorption features in the K band provide an
absolute wavelength and instrumental profile reference, similar
in concept to the iodine gas cell technique used in high-precision
optical RV exoplanet surveys (Butler et al. 1996).

We fit our observed spectra with the templates by fitting a number of
free parameters including a velocity shift and a power-law scaling factor
for both the stellar and telluric template.  The stellar rotational and
instrumental broadening are also free parameters, as is a second-order 
continuum normalization, and a second-order wavelength dispersion. The 
model spectrum is fit to each observed spectrum using
the Levenberg-Marquardt method (Bevington \& Robinson 1992) to optimize
the parameters of the model.  The wavelength shift of the stellar template
relative to the telluric template is then the measured RV of the star,
which we then correct for the motion of the barycenter.
We used a Monte Carlo technique to estimate the errors in
the model parameters. For each observation, we generated 100
simulated observations based on the measured noise in the spectrum and
refit the model to each of the simulated observations.
The standard deviation of the 100 results for each parameter, and in 
particular the RV, was taken as the statistical uncertainty for that
observation.

For the case of the Phoenix and IGRINS spectra where only the
one final observed
spectrum for each night is produced by the reduction (see \S 2.2.3 and
\S 2.2.4), the above procedure gives the final RV and associated uncertainty.
In the case of CSHELL and NIRSPEC data, we perform the above RV analysis
on each nod position in each exposure separately.  The 
final, nightly RV was then determined by calculating the average of
the individual nod RVs, weighted by their uncertainties. The
final uncertainty in the nightly RV was computed by taking the
weighted standard deviation of the nod RVs and dividing by the
square root of the number of nods.

The above procedure gives only an estimate of the statistical uncertainty
in each RV measurement based on the signal-to-noise in each spectrum.
We have previously assessed the long-term systematic uncertainties in our 
observations by routinely observing stars known to have stable RVs (i.e.
$<50~m~s^{-1}$; Prato et al. 2008; Mahmud et al. 2011; Crockett et al. 2012). 
For CSHELL we have determined that the long term stability in this technique
is 66 m s$^{-1}$ and for NIRSPEC we have determined this long term uncertainty
to be 30 m s$^{-1}$ (Crockett et al. 2012).  In the case of Phoenix, we have a
total of 9 observations of our RV standard GJ 281, and the standard deviation
of our RV measurements for this star is 62 m s$^{-1}$ which we take as the
systematic uncertainty for our Phoenix observations.  This value is very
similar to, but slightly better than what we get for CSHELL.  The slight
improvement may be the result of the fact that Phoenix has a somewhat higher 
resolution and about twice the wavelength coverage of CSHELL.  We have only
4 observations of GJ 281 with IGRINS, which is similar in resolution to
CSHELL but covers a wider wavelength range in the region of interest.  The
standard deviation of these observations is 52 m s$^{-1}$; however, given the
small number of observations, we assign a more conservative systematic 
uncertainty to our IGRINS measurements of 75 m s$^{-1}$.  We then add these 
systematic uncertainties in quadrature with the statistical uncertainty from 
each night's observation to obtain a final RV uncertainty for each observation.
Our measured RVs and uncertainties for CI Tau for each IR spectrograph are 
presented in Table 1.  

All of our K band RV measurements were calculated with respect to the same 
reference: the telluric spectrum.  Therefore, we combine the measurements from
all instruments together and analyze them as one group with no zero point
adjustments.  For the plots and
the values in Table 1, we have subtracted the mean of all the IR RV
measurements from the reported values.  We then used the Lomb-Scargle
periodogram technique (Horne \& Baliunas 1986) to search for periodicity
in the IR RV measurements.  Given the relatively large RV uncertainties
we obtain, our data are primarily sensitive to Jupiter mass or larger companions
in relatively tight orbits.  Therefore, our initial periodogram search was
in the range of 2 -- 20 days (the lower bound on the period set by the Nyquist
frequency since our observations are usually taken 1 day apart or longer).
The power spectrum is shown in Figure 1, where
a strong peak is apparent at 8.99 days.  We also searched the peridogram 
between 20 -- 100 days for completeness; however, no strong peaks appear in
this range (the periodogram peak in this range has a value of 5.8 with a 
false alarm probability of 0.22).  We estimated the false alarm
probability in the power spectrum peak of Figure 1 using a bootstrap method where
we randomly reordered our RV measurements to create new data sets 
observed with the same temporal sampling as our observations, ensuring
a consistent variance for each data set.  We then computed the
power spectrum of each data set over the same 2 -- 20 day period range as 
done for the original data and repeated this process 10,000 times.
In doing so, we found that the false alarm probability for the peak seen at
8.99 d in the IR data is $6 \times 10^{-4}$.  When we fit the RV data with a Keplerian orbit (below) 
and subtract this fit from the data, the 8.99 d peak and the nearby peak at 
$\sim 9.2$ d vanish from the power spectrum, indicating that there is only 
one potentially periodic signal present near this period.

Interpreting the periodic RV variation seen in the IR spectra of CI Tau
as orbital motion resulting from a low mass companion, we performed a Keplerian
fit to the velocity variations.  In the orbit fitting, we keep as fixed the 
orbital period as determined from the power spectrum analysis and treat
as free parameters the center-of-mass velocity of the system, the
eccentricity, the velocity amplitude for CI Tau, the longitude of
periastron, and the phase of periastron passage.  We use the nonlinear
least-squares technique of Marquardt (Bevington \& Robinson 1992) to find 
the best-fit parameters for the orbit 
(Table 4, column for IR only fit).  While we used the peak in the power 
spectrum as our initial guess for the period, we determined the more accurate
period, reported in Table 4, that
gives a $\chi^2$ minimum for the orbital fit by densely sampling periods near
the periodogram peak period and fiting a parabola to the resulting $\chi^2$
values.  The observed IR RVs for
CI Tau, phased to this 8.9965 day period, along with the orbital fit are 
shown in Figure 2;
there is considerable scatter around the fit (RMS of 0.694 km s$^{-1}$) which
we believe is astrophysical in origin.  We discuss the potential causes
of this in \S4.   Uncertainties in the orbital fit parameters are
derived by Monte Carlo simulation of the data: for 1000 simulations
we construct fake RV data using the RV fit and applying Gaussian random 
noise with a standard deviation equal to that in the residuals from the
fit of Figure 2.  We then analyze these model data using the same procedure
outlined above for the actual observations.  In doing so, we keep the
data uncertainties equal to the values reported in Table 1 for the purpose
of the orbit fitting.  The resulting uncertainties in the orbital parameters, reported as the standard
deviation of the derived properties,
appear in Table 4, and the distributions of the derived periods,
eccentricities, and $M_p{\rm sin}i$ values are shown in Figure 3.  The 
inclination dependent planetary mass of 8.81$\pm$1.71 M$_{Jup}$ was derived 
assuming a stellar mass of $0.80 \pm 0.02$ M$_\odot$ for CI Tau (Guilloteau 
et al. 2014).  The uncertainty of the planet's mass incorporates this 
uncertainty in the stellar mass.  Guilloteau et al. determined a circumstellar
disk inclination for CI Tau of 45.7$\pm$1.1$^{\circ}$; assuming that 
the planet and the disk are coplanar, we find an absolute mass for 
CI Tau b of 12.31$\pm$2.39 M$_{Jup}$.

\subsection{Optical RVs}

We followed the approach of Huerta et al. (2008) and Mahmud et al. (2011) 
and determined optical RVs using a cross-correlation analysis of nine orders
in the echelle spectra.  Each order contains $\sim100$ \AA, and the nine 
orders span the wavelength range $5,600 - 6,700$ \AA. Orders were chosen for 
analysis based on high signal-to-noise ratio, a lack of stellar emission 
lines, and a lack of strong telluric absorption lines present in the order. 
We used the mean of the RV measurements from the multiple echelle orders as 
the final RV value, while the standard deviation of the mean is assumed to be 
the internal statistical uncertainty in the RV measurement. We used the CI 
Tau observation with the highest signal-to-noise ratio (JD 2455160.819) as 
the template for the cross-correlation analysis.  Three of the observed 
optical echelle spectra were not suitable for RV determinations owing 
either contamination by a weak solar spectrum due to observations made through
moderate cloud relatively close to the Moon, or to 
relatively low signal-to-noise ratios in the continuum and/or large 
veiling values which resulted in weak photospheric lines and very large RV 
uncertainties; however, these observations remain useful for 
the emission line analysis.  The RVs for these data are not reported in 
Table 2.  The measured velocities were then corrected for
the motion of the barycenter and the mean RV of the optical measurements
was subtracted from all the values as we are only interested in relative
velocity variations in this study.  These final optical RV measurements are
presented in Table 2.  As discussed in previous studies from this series
(e.g., Huerta et al. 2008; Prato et al. 2008; Mahmud et al. 2011; Crockett
et al. 2012), we have observed several RV standards known to be stable at
a level of a few m s$^{-1}$ to assess the long term stability of our
measurement technique.  We find that our observations of these stars show
an RV standard deviation of $\sim 140$ m s$^{-1}$ for optical wavelengths, which we take as the
intrinsic uncertainty in our method.  We add this in quadrature to the
internal uncertainties determined above to get a final optical RV measurement
uncertainty, also reported in Table 2.

In \S4 we discuss the possibility that the RV variations recorded for
CI Tau result from an accretion hot spot.  This hypothesis can potentially
be tested by examining the variations in the veiling on CI Tau, and by
examining the behavior of the narrow component (NC) of emission lines
such as \ion{He}{1} 5876 \AA\ and the \ion{Ca}{2} IR triplet (IRT).
Veling in CTTSs is the apparent filling in or weakening of photospheric
absorption lines caused by a featureless continuum, believed to result 
from the shock which forms when accreting disk material impacts the
stellar surface (e.g., Hartigan et al. 1989).  In this study we are only
interested in the variations in the veiling, so we measure a veiling, $r$, 
relative to the observation of CI Tau with the strongest absorption lines.
An accurate measure of veiling on spectra with moderate signal-to-noise
such as these is aided by combining information from as many lines as
possible.  One such method of doing this is to use the least squares 
deconvolution (LSD) technique introduced by Donati et al. (1997).  This 
technique assumes the observed spectrum is the convolution of a single 
intrinsic photospheric line profile convolved with a series of delta functions
whose location and amplitude give the wavelength and intrinsic depth
of each line in the spectrum.  Using a constant line list, the LSD
technique can be used to deconvolve the spectrum to obtain the intrinsic
photospheric profile of each observation.  If the lines in a given 
spectrum all get weaker because of an increase in veiling, the recovered
intrinsic profile will also get weaker.  As a result, we can compare
the recovered LSD profiles from each observation and measure a very
accurate relative veiling using the observation with the strongest 
LSD profile as the reference spectrum.  For the measurements here, we
used the LSD code of Chen \& Johns--Krull (2013) and a custom line
list made using the VALD database (Kupka et al. 2000).  The final line
list contains a total of 1944 lines spanning the wavelength range
5350 -- 8940 \AA.  The majority of the lines are found in the 
5350 -- 6500 \AA\ range.  We used the observation from JD 2456605.906
as the reference spectrum, with veiling $r = 0.0$ by definition;
all other values of veiling reported in Table 2 are relative to this observation.

The \ion{He}{1} 5876 \AA\ and \ion{Ca}{2} emission lines of many CTTSs
appear to be made up of a narrow component (NC) sitting on top of a 
broad component (BC) of the line (e.g. Batalha et al. 1996; Beristain et al.
2001; Alencar \& Basri 2000).  Here, we are interested in the radial
velocity of the NC of the line.  In order to isolate just this component,
we follow the example of several earlier investigators (Johns \& Basri 
1995; Alencar \& Basri 2000; Sicilia--Aguilar 2015) and fit each observed
emission line with multiple Gaussians.  In our spectra, the only IRT line
present is the 8662 \AA\ line so we fit this and the \ion{He}{1} line.
Given the higher signal-to-noise ratio and complexity of the former, we required
four Gaussian components to properly fit the \ion{Ca}{2}
line while we needed just two Gaussian components to fit the \ion{He}{1}
line.  For each spectrum we averaged the wavelength solution from the
Th-Ar lamp spectra taken immediately before and after each stellar 
spectrum.  The wavelength of the NC fit was then translated into an RV
and the motion of the barycenter was removed.  We estimate the
uncertainty in the RV by performing a Monte Carlo analysis
on the fit, adding in normally distributed random noise at a level 
given by the signal-to-noise in each observation for the line in question.
A total of 100 Monte Carlo trials were performed for each line fit and
the standard deviation of the resulting RV values was taken as the 
uncertainty.  To this, we added in quadrature the 140 m s$^{-1}$ systematic
uncertainty identified above for the photospheric RV analysis.  The NC RV
and its uncertainty for both the \ion{He}{1} 5876 \AA\ and 
\ion{Ca}{2} 8662 \AA\ lines are reported in Table 2.

CI Tau was identified as a potential host of a several Jupiter mass
planetary companion by Crockett et al. (2012) because of its significant 
optical RV variations and IR (K band) RV variations of similar
amplitude; however, these results were based on a relatively small number of 
data points (10 optical and 14 IR).  We now have 26 optical RV data points.  
We computed a power spectrum of the optical RV measurements, again using the 
Lomb-Scargle periodogram technique (Horne \& Baliunas 1986).  The strongest 
peak in the power spectrum occurred at a period of $\sim 9.5$ d with a false 
alarm probability (determined from a Monte Carlo analysis) of 0.017.  The 
second strongest peak in the power spectrum is nearly as strong and occurred 
at a period of $\sim 7.2$ d, also with a false alarm probability of 0.017.  
These peaks are suggestive, particularly since the 7.2 d period is very close 
to the period found below for the variability of the optical photometry 
(\S3.3), and the 9.5 d period is close to the 8.99 d period found for the 
variability of the IR RV measurements above (\S3.1).  However, to 
independently confirm {\it periodicity}, it is generally desirable to have a 
false alarm probability that is lower than the 0.017 observed in the
optical data.  As a
result, the optical data alone do not represent as significant a detection of a periodic signal
as the IR data.

\subsection{Combined IR and Optical RVs}

Given the periodic signals suggestive of a giant exoplanet with a period of
$\sim$9 d (IR) to $\sim$9.5 d (optical), we combined the
optical and IR mean subtracted RV data into one time series for analysis.
We again used the Lomb-Scargle periodogram technique to compute the power
spectrum of this combined data set
(Figure 4).  The peak near 9 days has become even stronger, but has shifted
slightly to 8.9965 d instead of 8.9891 d; however, this is well within the
period uncertainty determined from the analysis of the IR RV data alone
(Table 4).  We again use a bootstrap Monte Carlo technique to estimate
the false alarm probability for this peak, this time utilizing 10$^6$ trials
sampling the period range 2 to 20 days as done initially for the analysis
of the real data.  From this simulation, we estimate a false alarm 
probability of $8 \times 10^{-6}$.  We also performed a periodogram 
analysis of the data in the range 2 to 100 days and find that the 8.99
day period remains the strongest peak in the data.  The Monte Carlo 
simulation sampling the same 2 to 100 day period range to estimate the
false alarm probability again yields a value of $8 \times 10^{-6}$.

We follow the same procedure described above to fit a Keplerian orbit to this combined
dataset and to estimate the uncertainties in the orbital parameters.  The
RV fit is shown in the bottom panel of Figure 4, which is almost identical
to Figure 2.  This suggests that the optical data
do show evidence for the planetary companion, but that activity-induced
RV noise muddies the planet's signal in the optical data to some degree.
The RMS of the fit residuals to the combined IR$+$optical RVs
is slightly greater at 0.728 km s$^{-1}$ than for the IR data alone, again 
likely the result of
the increased activity related noise in the optical RV measurements.  The orbital parameters 
and associated uncertainties are given in Table 4 and the distribution of
the derived periods, eccentricities, and $M_P {\rm sin} i$ are shown in
Figure 5. We have subtracted this RV fit from the RV data points and 
recomputed the power spectrum of the residuals.  The highest peak in the
power spectrum occurs at a period of 4.75 d and has a false alarm 
probability of 0.07.  As an additional test to see how strongly the optical
data might be affected by (and potentially biased by) spot-induced RV noise,
we subtracted the IR only RV fit (Figure 2) from the optical RV data and
computed the power spectrum of those residuals.  Again, no significant peaks
were found: the strongest occurs at $\sim 4.6$ d with
a false alarm probability of 0.146.  This could indicate that the 
activity-induced variations of CI Tau do not remain coherent over the
$\sim 9$ year span of this data, the result perhaps of the migration of spots
(e.g., Llama et al. 2012) or their disappearance from one region of the stellar surface
and reemergence in another over the years of observation.  Because CI Tau
is a CTTS, there are many potential sources of variation in addition to dark,
cool spots.

The IR data alone phase nearly as well to the 8.9891 d period, found for the
combined IR plus optical data set (Figure 4), as it does to the 8.9965 d IR only period (Figure 2).
The optical data in the bottom panel of Figure 4 phases fairly
well to this period, although there are a few significant outliers that may
represent times when spot induced noise was particularly problematic.
We emphasize that the amplitude of the optical RV variations is the
same as that of the IR RV variations to within their uncertainties.  If we
hold the orbital period and eccentricity fixed to the values found from 
the combined fit and fit only the IR RV data points we find a velocity 
amplitude of $K = 0.99 \pm 0.17$ km s$^{-1}$.  If we then fit only the
optical RV points holding the period and eccentricity fixed to the same
values, we find $K = 0.63 \pm 0.23$ km s$^{-1}$ (the lower significance
of this fit is a combination of the fewer number of optical observations and
the additional scatter present in the optical RVs, which in turn prevented a
definite detection of this signal in the optical-only periodogram analysis
described above).  Thus, the optical 
amplitude is found to be lower than that in the IR, but the difference is
not statistically significant.  

The ratio of the optical to IR RV amplitudes is $0.64 \pm 0.26$.
If cool, dark spots were responsible
for the RV signals in both the optical and IR, we would expect this ratio of
the optical to IR RV amplitudes to be $>$4 because spot noise has a higher 
impact on the optical RVs (Mahmud et al. 2011).  From the combined
dataset, the candidate planet's mass is 
$M_P {\rm sin} i = 8.08\pm1.53$ M$_{Jup}$.  Again
assuming a stellar mass of $0.80 \pm 0.02$ M$_\odot$ for CI Tau (Guilloteau 
et al. 2014), and using an inclination of 45.7$\pm$1.1$^{\circ}$ for
CI Tau b based on the inclination of the circumstellar disk (Guilloteau et al.),
we get an absolute mass for CI Tau b of 11.29$\pm$2.13 M$_{Jup}$.
These values are very similar to those determined using the IR RV measurements alone.

The eccentricity we find for CI Tau with the combined IR and optical 
orbital solution, 0.28 $\pm$0.16, is relatively large compared with typical 
eccentricities of hot Jupiters orbiting mature stars, which are usually $<$0.1 
(Fabrycky \& Tremaine 2007).  However, the eccentricity of a 
few Myr old object is likely to be a property which either evolves toward zero 
as the result of dynamical interactions with disk material and/or other 
planets, or could potentially be a property which dooms a massive planet to 
orbital decay and consumption by its parent star.  Alternatively, 
high-eccentricity giant planets may be dynamically ejected from their
host system.  Anecdotally, with the accumulation of larger data sets, hot 
exo-Jupiter eccentricity estimates tend to decrease.  Thus as we collect more 
data we will examine the cumulative changes, if any, in eccentricity and 
other orbital parameters.

\subsection{Optical Light Curve}

We used both a Fourier-fitting routine (Harris et al. 1989) and the 
Lomb-Scargle method (Horne \& Baliunas 1986) to look for photometric
periodicity.  Both approaches yielded significant power at a period of 
$\sim 7.1$ days.  The latter technique gives a false-alarm probability of 
$<10^{-4}$.  This value was again obtained by running 10,000 Monte Carlo 
simulations of the data, sampling the observed photometry randomly over the 
epochs of observation.  Figure 6 displays the Lomb-Scargle power spectrum. 
While the power in the periodogram is quite strong, and the data clearly 
show systematic variations, the periodogram only samples this potential period in a
limited way.  The photometric data were taken in 3 runs spanning a total
of 34 days.  Figure 7 shows the photometry with each observing run color
coded.  The top panel shows the data phased to the 7.1 d period found in
the periodogram analysis.  The bottom panel shows the data phased to the
8.99 d period found in the RV analysis (the figure looks the same whether
we phase to 8.997 d or 8.994 d as these two periods are so close and the
length of the photometric campaign was so short that there is a maximum
phase shift of only 0.002 between the two periods).  Clearly the data
are not strictly periodic in either panel, though in the top panel (7.1 d)
the data from all 3 runs show a decline in brightness at approximately 
the same phase.  For the two runs that cover the latter phases, the
brightness recovers fully at about the same phase as well, but one of
these runs shows a substantially deeper minimum.  Whatever is causing
the decline in brightness changed measureably from one phase to the next,
or possibly other factors contributed to augment the dimming.
The bottom panel, phased to 8.99 d, shows no clear behavior from one
phase to the next.  Because of the intrinsic jitter in the variability of 
this system, we are not able to determine a definite photometric period
for CI Tau; however, the data do not support a period near 9 days, and
instead point to a period closer to 7 days for this star.

\subsection{H$\alpha$ Analysis}

All 29 optical echelle spectra of CI Tau show strong, variable
H$\alpha$ emission.  Figure 8
shows the average H$\alpha$ profile of CI Tau plotted in the stellar rest
frame.  This average profile has an emission equivalent width of 69.6 \AA.  
We computed the power spectra, again using the Lomb-Scargle method
(Horne \& Baliunas 1986), of the relative flux variations in
each 5 km s$^{-1}$ velocity channel across the H$\alpha$ emission line.  
We found that the velocity channel at $\sim$200 km s$^{-1}$ (Figure 9a) shows 
the strongest power, 11.0, in the periodogram analysis; this peak occurs at a
period of 9.4 days.  The next two strongest peaks in this channel appear at
periods of 9.0 and 9.2 days.  The surrounding nine independent velocity 
channels, between 181 and 227 km s$^{-1}$, also show a power spectrum peak in 
their relative H$\alpha$ flux variations at 9.4 days.  Figures 9b and 9c
illustrate the power spectrum from two other velocity channels in the 
H$\alpha$ line.  Figure 9b shows the power spectrum of the 0 velocity channel,
while Figure 9c shows the velocity channel at $-135$ km s$^{-1}$ which is
the channel that shows the strongest fractional variation in the profile.
This blue-shifted velocity channel is near the center of a variable absorption
component that sometimes appears the profile, indicative of a variable wind
flowing from CI Tau.  These additional power spectra indicate the ``typical"
strength of the periodograms outside the strongest one at $\sim 200$ 
km s$^{-1}$.  The flux variations for the the 200 km 
s$^{-1}$, phased to a 9.4 day period, are shown in Figure 10a.
Again using a Monte Carlo analysis to resample
the observed flux values in a random order while preserving the actual dates
of observation, we found that the 9.4 day period  peak in the 200 km s$^{-1}$
velocity channel has a false alarm probability less
than $10^{-4}$ and any peak stronger than 9.8 has a false alarm probability of
$10^{-3}$ or less.

While the phased flux curve in Figure 10a shows little scatter, the power 
spectrum in Figure 9a shows strong power at many periods, making it unclear 
whether
there is a true period present.  If we fit the phased flux curve in Figure 10a 
with a sine wave and subtract the fit from the data, we can compute the
power spectrum of the residuals.  Doing so gives a periodogram with a peak
of only 6.3 at a period of $\sim 2.3$ d, near the theoretical Nyquist limit
for data sampled 1 day apart as is typically the case for the individual
runs on which these observations were obtained.  The false alarm probability
for a peak this strong is 0.254.  For periods longer than 3 d,
the peak in the residual power spectrum occurs at $\sim 14.7$ d with a power
level of 5.5, corresponding to a false alarm probability of 0.499.  While the 
peak in the power spectrum of the flux variations occurs at $\sim 9.4$ d, the 
next two peaks are very close in strength and produce phased variability with 
little scatter.  For example, phasing to the
peak at $\sim 9.0$ d produces the curve shown in Figure 10b.  Thus we conclude
that if there is periodic modulation in the H$\alpha$ line of CI Tau, there is 
only one significant period in the 9.0 -- 9.4 d range.  These peak periods are 
suggestively 
close to the same period as found in the IR RV variations, possibly indicating
that the source of the RV variations is also influening the behavior of the 
H$\alpha$ line.

\section{Discussion}

\subsection{The Role of Accretion Hot Spots}

Dark, cool spots can produce RV signals on a rotating star
that mimic those from a low mass companion (e.g., Saar \& Donahue 1997;
Desort et al. 2007; Reiners et al. 2010).  To first order, this results 
because the dark spot removes a contribution to the photospheric absorption
lines at the projected RV of the region of the rotating star 
where the spot is found.  This leads to a distortion in the line profile 
which can appear as a small RV shift.  
In the case of dark spots, their presence can be diagnosed on the basis of 
the wavelength dependence of their effect;
spots are not completely dark, and become much less so at IR
wavelengths relative to the optical (e.g., Martin et al. 2006; Huelamo et al.
2008; Prato et al. 2008; Mahmud et al. 2011).  This fact, and the observation
that the RV amplitude of CI Tau is nearly identical in the optical and the
IR, suggests that dark cool spots are not the cause of the
observed RV signal in this star; however, there remains the possibility of
hot spots.

Bright accretion spots on classical T Tauri stars can in principle create
the same apparent RV signal as dark spots (e.g., K\'osp\'al et al. 2014;
Sicilia-Aguilar et al. 2015).  Accretion hot spots which produce veiling
on CTTSs typically have temperatures $\sim 10,000$ K and produce a largely
featureless continuum (Hartigan et al. 1989; Basri \& Batalha 1990; Hartigan
et al. 1991; Valenti et al. 1993).  As a result, these hot spots do not 
contribute to the cool ($\sim 4,000$ K) photospheric absorption lines at the 
projected RV of the spot and distort the line profile shape
in the same way that a dark, cool spot would.  Furthermore, because these 
spots are hotter than the stellar photosphere, it is expected that they
produce essentially identical line profile distortions (and hence apparent
RV signals) in the optical and the IR.  Thus, we can not
use the similarity of amplitude in the optical and IR RV signals to rule out
bright accretion spots as the source of the observed RV signals in CI Tau.
However, if bright accretion spots are responsible for the 
observed RV signals, various simple predictions may be explored to test
this hypothesis.  These include 
photometric variability produced by the accretion spots, potential correlation
between the veiling and the measured RV signals, and anti-correlation between
the RV of lines formed in the hot spot with the photospheric
RV signals.  We consider each of these in turn.

The IR RV measurements presented here were taken over a time span of $\sim 5$
years.  Including all the optical data, the time span over which all the
RV measurements we obtained is $\sim 10$ years.  The RV measurements 
appear to be well phased over the 5 years in which the IR observations 
were made (Figure 2), and with a small modification to the period (well 
within the period uncertainty), the full dataset shows good coherence over 
10 years (Figure 4).  In order for a hot spot to produce such a RV signal,
it too would have to be coherent over a similarly long time span, and thus
might be expected to show rotationally modulated photometric behavior.
Our own photometry presented above does show apparent modulation; however,
while not well-determined, the period is not consistent with the 8.99 d
period found in the RV signals (Figure 7).  In addition to our own 
observations of CI Tau, others have monitored this star photometrically.  
Grankin et al. (2007) observed CI Tau photometrically a total of 320 times 
between 1987 and 2003.  Artemenko et al. (2012) report a period of 16.10 days
based on this data.  We have downloaded this photometric database and performed 
periodogram analyses on the entire dataset and on each observing season
subset of the data.  The highest peak we recover in the power spectrum of
the entire dataset is at a period of 16.24 days with a false alarm probability
of 4.6\%.  We assume this is the same signal Artemenko et al. report at a
period of 16.10 days.  We do not consider this a firm detection.  Analyzing
each observing season individually does not reveal stronger periodicity, and
in no case do we find significant periodicity near
9 days.  Percy et al. (2010) used a ``self-correlation" analysis on this
same photometric data, augmented by a few (12) additional observations and
report a period of 14.0 days for CI Tau.  While the photometric period for
CI Tau remains uncertain, the existing photometry does not suggest a period
of 9 days for this star.

In addition to producing a photometric signal on CI Tau, if a hot spot
is the cause of the observed RV variations, it is possible that
there will be a relationship between the observed RV signals of CI Tau and
the veiling.  When the hot spot is most directly facing the observer (effectively
in the middle of the star), the veiling should be largest and the RV should
not be affected.  As the spot moves to either limb, the veiling should
decrease and the RV will become either red- or blue-shifted depending on
which limb the spot is on.  Thus, a plot of veiling versus photospheric
RV might show a parabolic relationship with the veiling largest at 0 
relative velocity.  For our optical data, the observed relationship between
veiling and photospheric RV is shown in the bottom panel of Figure 11;
no correlation was observed.  We performed a correlation analysis
(both linear correlation and Spearman's and Kendal's $\tau$ rank order
correlation) on the veiling and optical photospheric RV measurements and 
found no relationship at all between the two quantities.

The above predictions relating photometric brightness or veiling to the
photospheric RV caused by a hot spot implicitly assume that when the
accretion related continuum emission changes intensity, the property of the 
hot spot that is primarily varying is its projected area.
If on the other hand, accretion variability causes the surface 
flux from the hot spot to change substantially on short timescales, it
may be difficult to detect rotationally modulated signals from the hot spot
related continuum emission.  Spectroscopic studies of the veiling continuum
emission (e.g. Valenti et al. 1993; Gullbring et al. 1998; Cauley et al.
2012) find that the accretion continuum is produced in marginally optically
thick gas.  These studies also find that the temperature of the gas producing
the hot spots are all close $\sim 10,000$ K.  This suggests that the surface
flux of the accretion hot spots is very similar from star to star.  Using
the parmeters of the accretion emission published by Cauley et al. (2012),
we find that the accretion luminosity is very well correlated with the 
area of the star covered by the accretion spots.  The studies above compare
the properties of accretion spots from one CTTS to another; however,
simulations of variable accretion onto individual CTTSs show that variations
in the area of the accretion spots are highly correlated with the instantaneous
accretion rate (e.g. Romanova et al. 2008; Kulkarni \& Romanova 2008; 
Kurosawa \& 
Romanova 2013).  Indeed, Batalha et al. (2002) performed a variability study
of the CTTS TW Hya and find that the projected area of the hot spot is
well correlated with the hot spot luminosity and resulting veiling.  
Therefore, we suggest that the hot spot emission strength on CI Tau may
be a good diagnostic of the projected area of the hot spot on this star
as assumed in the tests described above.

Another prediction of the accretion hot spot on a rotating star hypothesis is
that there should be apparent rotational modulation of emission lines formed
in the accretion footprints themselves.  Further, because these emission
lines reveal the actual motion of the hot spot, whereas the impact of the hot
spot on the photospheric absorption lines is to remove a contribution to
the line profile, the modulation of the hot spot emission lines should be
180$^\circ$ out of phase with the photospheric lines.  Thus, plotting the
emission line RV versus the photospheric RV should show an inverse 
correlation.  Such behavior has been seen in the star EX Lup by K\'osp\'al
et al. (2014) and has been modeled as an accretion spot by Sicilia--Aguilar
et al. (2015).  The lines showing this behavior are the narrow components (NCs)
of metallic emission lines such as that from \ion{He}{1} 5876 \AA\ and the
\ion{Ca}{2} IR triplet.  It has long been expected that the NCs of
these and other emission lines form in the post-shock region (e.g.
Batalha et al. 1996; Beristain et al. 2001) at the base of the accretion
footprints.  As a result, the NC of these emission lines serve as a good
indicator of the behavior and location of accretion footprints, and
they have even been used to Doppler image the location of accretion footprints
on CTTSs (e.g., Donati et al. 2008, 2010).

We have estimated the location and size of the hot spot required to produce the
observed photospheric RV modulation using the disk integration code 
utilized by Chen \& Johns-Krull (2013).  A single hot spot is assumed on
the surface of CI Tau, and we assume a stellar inclination of 45.7$^\circ$
(Guilloteau et al. 2014) and a $v$sin$i$ = 11 km s$^{-1}$ (Basri \&
Batalha 1990).  We find a best fit to the observed photospheric RV
measurements for a hot spot located at a latitude of 82$^\circ$ covering a
maximum projected area 46.7\% on the surface of the star.  Such a large
areal hot spot coverage is far greater than values typically found on CTTSs
which are usually in the few percent range (e.g., Valenti et al. 1993;
Calvet \& Gullbring 1998).  However, there is a degeneracy between the
spot latitude and the size such that a hot spot covering only 10\% of the
stellar surface at a latitude of 16$^\circ$ produces an almost equally good fit ($\chi^2$ reduced by
only 4\%).  Such a relatively small, low latitude hot spot would show a
large RV modulation (10 km s$^{-1}$ peak to peak) that is 
180$^\circ$ out of phase with the photospheric RV measurements, producing 
an inverse correlation between the photospheric RV measurements and RV
measurements for emission lines coming from the hot spot.  We looked for this 
behavior in the NCs of the \ion{He}{1} 5876 \AA\ and \ion{Ca}{2} 8662 \AA\
lines.  

The RV signals of these two emission lines are plotted versus the optical 
photospheric RV measurements in the top two panels of Figure 11.  
We performed correlation analyses using 
both the linear correlation coefficient and the Spearman's and Kendal's
$\tau$ rank order correlation coefficients.  No correlation was observed.  The most
significant correlation with the photospheric RVs is for the RV measurements
of the \ion{He}{1} line, but the false alarm probability is 0.58 for 
Kendal's $\tau$ and higher still for the other statistics.  We also
performed a periodogram analysis on the emission line RV measurements
and the veiling measurements, as well as phase folding these to the 
8.9965 and 8.9891 d periods.  
In all cases, no significant signal was
found.  As a result, there is no evidence that a hot spot is 
producing the RV signals seen in the photospheric absorption lines, and
long term photometric measurements do not show the signal expected from
a long lived coherent hot spot if it were responsible for the observed
RV variations in CI Tau.  Thus, we suggest that it is unlikely that
a hot spot is responsible for the photospheric RV signals seen in this star.

\subsection{Scattering off an Inner Disk Wall}

CI Tau is a CTTS surrounded by a circumstellar accretion disk.  The
disk mass for CI Tau has been estimated by several authors 
\citep{and05, and07, moh13, mcc13} with values that range from 18.7
M$_{Jup}$ \citep{moh13} to 71.3 M$_{Jup}$ \citep{mcc13}.  \citet{mcc13}
estimate that the inner disk of CI Tau is truncated at a radius of
0.12 AU, which is close to the apastron distance (0.10 AU assuming $e = 0.28$)
of the suspected
planet found in our RV analysis.  The inner disk can scatter 
incident starlight, adding a scattered light spectrum to the directly
observed spectrum of the star.  Such a scattered light spectrum was
detected in the optical for the spectroscopic binary star KH 15D by
\citet{her08}.  These authors find that the reflected light from the
disk is about 3\% of the direct spectrum in the optical near 6000 \AA, 
and that the reflected component can cause measurable effects on optical
line profiles at certain phases in the binary orbit.  For a single star
with an azimuthally symmetric disk, this type of scattering should produce
a symmetric reflected light line profile centered on the stellar line,
and would not therefore be expected to produce any apparent RV
shift.  However, the inner walls of circumstellar disks are believed
to be warped or otherwise not symmetric in many cases, resulting in
detectable photometric variability \citep[e.g.][]{cod14, sta14, mcg15}.
If there is some coherent structure at or near the inner
wall of the accretion disk, it might contribute a scattered
light component that is Doppler shifted along our line of sight relative to
the star as a result of the orbital motion of the disk.  As the structure
orbits the star, its velocity shift relative to the star would change,
potentially distorting the photospheric absorption lines and creating an 
apparent velocity signal for the star.

For a structure at the inner wall of the disk to be responsible for the
observed RV signatures above, it must be located at a
distance where the period is equal to 8.99 days.  For the 0.80 M$_\odot$
mass of CI Tau, this corresponds to 0.079 AU, well inside 
the 0.12 AU inner wall of the disk found by \citet{mcc13}.  Looking at it
another way, the orbital period at 0.12 AU is 17 days, 
substantially longer than the period of RV variations.  In order
to produce the observed RV variations, such a disk structure would need
to remain stable over the 9 year span of the data collected here.
Such stability is unlikely given the short dynamical time of the disk
at this radius, unless the disk structure, such as a warp, is excited 
and maintained by some other object or process.  A massive planet inside
the disk gap could excite such a disk structure.
The interaction of the
inner disk wall with the stellar magnetosphere, particularly if the 
magnetosphere is tilted, could also excite a long term stable disk warp as is
believed to be the case for AA Tau for example \citep{bou03, bou13}. 
Such a disk warp is tied to the stellar rotation, which our photometric
monitoring suggests has a period of $\sim 7$ days instead of
9 days.
Another estimate for the rotation period of CI Tau can be made using its
$v$sin$i$, stellar radius, and inclination.  Assuming the inclination of
the star is equal to that of the disk, we can use $i = 55^\circ$ and
$R_* = 1.41$ R$_\odot$ \citep[both from][]{mcc13}.  Taking $v$sin$i = 11$ km
s$^{-1}$ again from Basri and Batalha (1990), the rotation period is estimated
to be 5.3 days.  This estimate shrinks somewhat to 4.6 days using the 
inclination of 45.$^\circ$7 degrees from \citet{gui14}, and $i = 90$ gives an 
estimated period of 6.5 days.  There are of course uncertainties in the stellar
radius and $v$sin$i$, so these period estimates are themselves uncertain;
however, the available measurements point to a rotation period noticeably
less than the 8.99 day period of the RV variations.

While the estimates above argue that some sort of structure
in the inner disk wall scattering starlight from CI Tau is probably not
responsible for the RV variations we measure, we can look to the data for
evidence one way or another.  The spectral model used to measure the
K band RVs in \S 3.1 includes a term that measures the strength of the 
2.29 $\mu$m K band
photospheric lines which can vary as the result of veiling, a meaure of the continuum 
emission from the inner disk relative to that from the star 
\citep[e.g.][]{fol99, joh01}.  If there is a coherent structure in the
inner disk responsible for scattering starlight and producing the observed
RV variations, we might expect there to be a correlation between the K band
veiling and the measured RVs.  We convert our measurements into the 
K band veiling (measured at 2.29 $\mu$m as opposed to averaged over the entire
band), referenced to the model spectrum used in the fitting process.
These K band veilings are
reported in Table 1.  We phase the K band veilings with respect to
the 8.99 day RV period in Figure 12, but we find no apparent pattern with
the phase.
%We then computed the linear correlation
%coefficient between these veiling measurements and the RV values, finding
%$r = -0.06$ with an associated false alarm probability of 0.63.
We also
computed the periodogram of the K band veiling measurements, finding no
significant peaks near 9 days.  In particular, the false alarm probability 
of the highest
peak within $\pm 5$ days of 8.99 days is 0.81, with the period of the
peak being 13.6 days.  Thus, our IR observations
do not show any relationship between the disk emission and the measured
RVs.

We can also estimate the level of scattering that
would be needed to produce the observed RV signal.  As mentioned above, in
order to produce periodicity at $\sim 9$ days, the scattering surface in
the disk would need to be at radius of $\sim$0.08 AU where the Keplerian velocity
is $\sim 96$ km s$^{-1}$.  Using $i = 45.^\circ$7, this surface would vary
in RV by $\pm 68$ km s$^{-1}$ relative to the star.  As the 
lines in the scattered light spectrum move through the stellar spectrum,
they can distort the photospheric line profile and appear to produce a 
velocity shift.  We estimate the strength of the required scattered
light spectrum by asking how strong the scattered light line profile 
would need to be relative to the stellar lines in order to produce the
$\pm 1$ km s$^{-1}$ shift that is observed in the RV signal (Figure 4).
We first assumed the scattered light spectral lines are identical to 
those in the star.  This assumes the scattering source is essentially a point
in the disk so that the reflected stellar spectrum in not smeared in velocity
space because of formation around a range in azimuth in the disk.  We used the 
the LSD Stokes I photospheric profiles calculated for the optical spectra
in \S 3.2 and added a scaled version of the same spectrum to the original
profile shifted by a specified RV value.  We stepped through shifts of 
$\pm 68$ km s$^{-1}$ in 0.1 km s$^{-1}$ steps and used the same cross
correlation technique as in \S 3.2 to measure the resulting RV 
shift.  We find that in order to produce a maximum line distortion of 
$\pm 1$ km s$^{-1}$, the scattered light spectrum needs to be $\sim 17$\% 
the strength of the directly observed stellar spectrum (Figure 13).  If 
this were indeed occuring, a feature 17\% as deep as the 
observed profiles would be present in the spectra of CI Tau and would move
periodically back and forth relative to the main lines by $\pm 68$ km
s$^{-1}$.  Such a feature would be obvious in the LSD optical profiles (Figure
13) and is not seen.  If we instead represent the scattered light spectrum as 
the stellar spectrum convolved with a Gaussian of FWHM = 40 km s$^{-1}$ to
mimic scattering from a range of disk azimuth, we find that the scattered
light component must be $\sim 57$\% as strong as the stellar spectrum in
order to produce the measured RV variations.  Such a strong component
is not seen in the observed spectra, and this high level of scattering is also
unphysical \citep[e.g.][]{whi92}.  We conclude
it is highly unlikely that scattering off the inner disk is producing the
RV variations we observe.

\subsection{The Challenge and Necessity of Finding Planets Around CTTSs}

We conclude that the best interpretation of the data presented in this
paper is that there is an $\sim11-12$ M$_{JUP}$ mass planet on a somewhat
eccentric $\sim$9 d orbit around the CTTS CI Tau.  We have
illustrated some of the unique difficulties in searching
for planets around actively accreting, young stars.  However, in order to understand
planet formation, it will be necessary to look for planets in just such
systems.  All indications of the thousands of planets identified by the 
{\it Kepler} mission, as well as the structure and likely dynamic history of
our own solar system, point to significant evolution and migration of 
planets, including the gas giants (e.g., Levison et al. 2007).  In order to 
begin to document and characterize the extent of these processes, and to 
determine the nature of planet formation itself, we are compelled to search 
for the first generation of planets around host stars with extreme and variable properties.

Classical T Tauri stars undergoing active accretion present complex challenges
for the identification of even giant planets on short period orbits.  Some
of these have been described above.  Potential sources of variable photometry
include changes in geometry and extinction resulting from modifications in the
line of sight across or through the circumstellar disk as it rotates.  
Furthermore, bright accretion footprints and episodic accretion events, cool
stellar spots, stellar flares, massive coronal mass ejections, stellar jet 
outbursts, among other phenomena, may all contribute to a particularly high 
level of activity and thus variability.  Spectral  absorption line profile 
variability may result from the shift in an absorption line center as a large 
dark spot (or spots) is carried across the observed stellar hemisphere by the
star's own rotation.  Hot spots from accretion or flares on the stellar 
surface may produce an analogous result.  Accreting hot gas, and warm dust 
grains in the inner disk, may give rise to a continuum excess which veils 
absorption lines and can, in extreme cases, effectively obliterate them.
Strong variability in emission line fluxes and line profiles result from 
excitation arising in clumpy accretion flows, stellar winds, and jets (e.g.,
Alencar et al. 2005).  Yet it is at this tumultuous phase in a star's 
lifetime during which planets must have already formed or be in the formation
process, given the relatively short window of availability of the reservoir 
of raw material in the primordial disk.  

The evidence we present for a giant planet in the CI Tau system
is demonstrated on the basis of diverse data sets collected over 10 years at 
5 different facilities.  Over this multi-year time scale we find consistent 
variability in the IR RVs, as well as evidence in the optical RV variations, for 
this same periodic signal, supporting our planetary companion interpretation.
Some significant scatter is obvious in the IR RVs plotted in Figures 2 and 4; 
we interpret this as likely the result of astrophysical processes, such as those
described above.  Although working in the IR diminishes the impact of cool
star spots and stellar activity, it does not guarantee immunity from these 
phenomena, particularly in a classical T Tauri system.  Some processes, such 
as emission from warm grains in an inner circumstellar disk, may wield a 
greater impact in the IR.  However, although this work is outside the scope 
of this current paper, we are hopeful that it will be possible to minimize 
these sources of interference in our RV measurements. For example, for stars
with known rotation periods, specific observing allocations in the future can
be used to tailor the IR
spectroscopic observations to take place repeatedly at the same rotational 
phase of the star (e.g., Robertson et al. 2015), thus nulling any spot signal.
We can also experiment with spot modeling in order to determine the degree of
interference anticipated in our IR RV measurements, and with extracting 
contaminating spot signals from our spectra directly (e.g., Moulds et al. 
2012; Llama et al. 2012; Bradshaw \& Hartigan 2014; Aigrain et al. 2015).

The behavior of activity on young stars is not well understood.  Astronomers 
know that extreme cases are possible, for example unusually long lived spots 
which appear to phase coherently over many years (Stelzer et al. 2003; Mahmud
et al. 2011; Bradshaw \& Hartigan 2014), or spots with filling factors that 
cover most of the stellar surface (Hatzes 1995).  Although T Tauri stars are
all presumed to have strong magnetic fields and corresponding activity 
(Johns--Krull 2007), it is not uncommon to find systems which defy 
characterization of their rotation periods on the basis of spots (e.g., Xiao
et al. 2012).  Classical T Tauri stars, with their many types of potential 
activity, are among the most difficult for the measurement of rotation, 
although many show variable behavior which at times reveals periodic light
curve behavior (Herbst et al. 2002).
CI Tau appears to fit into this category.  The peculiar activity that 
distinguishes CTTSs is embodied by the H$\alpha$ line and its variations.
This line is a strong accretion diagnostic, and the H$\alpha$ variability of 
CI Tau is intriguing.  The red side of the profile shows potentially
periodic variability that phases well with the orbital period of the likely
planetary companion (Figure 8).  The exact cause of this variability is
not clear however.  The H$\alpha$ luminosity of accreting roughly planetary
mass objects may be as high as 10\% that of the central star 
(e.g., Zhou et al. 2014), so it is possible this variability results from
the RV motion of H$\alpha$ emission associated with the planet itself.
In that case, one would expect similar periodicity at all velocities sampled
by the planetary orbit, as long as they are strong enough relative to the
line profile variations from the star itself.  It is also possible that this
apparent periodicity in the H$\alpha$ line is caused by the planet modulating
the accretion of disk material onto the star, similar to that seen in close,
eccentric binary CTTSs like DQ Tau (e.g., Basri et al. 1997).  In either case,
the H$\alpha$ variability of CI Tau deserves further investigation.

The definitive characterization of a massive planet in the CI Tau system will 
require continued monitoring for an RV signal in the IR consistent with 
the orbital parameters identified to date and corroborating results from 
optical RV observations, modulo potentially variable cool spot noise.  A firm
detection of the photometric rotation period with a period different from
the RV period would also help substantiate the existence of a planet
orbiting CI Tau.  Verification of our result must necessarily rely for now
on these ground-based techniques as the candidate CI Tau 
planet will not be astrometrically detectable by the GAIA mission, for example
(Sozzetti et al. 2014).

\section{Summary}

We have identified a $\sim 9$ day period in the K band RV variations of the
classical T Tauri star CI Tau.  The best interpretation of these data is that
a massive planet is in orbit around this young star
located in the Taurus star forming region.  
This identification is based on high-resolution IR spectroscopy supported
by high-resolution optical spectroscopy and optical photometry, all collected over
a 10 year total time span.  For the $\sim 5$ years of our IR 
observations of CI Tau, from 2009 to 2014, the RVs extracted from the 
spectroscopy phase to a period of $\sim 9$ days.  While not sufficient
to independently confirm the $\sim 9$ day period, the full set of 
optical RVs also phase reasonably well to the $\sim$9 day RV period observed
in the IR spectroscopy.  While it is expected that the optical RV measurements
experience a greater impact from cool starspots than the IR RVs, the amplitude of the optical
RV variations is very similar to those observed in the IR, indicating that
noise from cool spots does not obliterate the planetary signal.  We also investigate the
possibility that the observed photospheric RV variations on this CTTS
result from an accretion hot spot or from scattering off the inner wall
of the accretion disk.  These scenarios gives rise to a few potentially
testable predictions which are not supported by the data collected here
or by other investigators.  Therefore, we find that the best interpretation
of the observations presented here is that the RV variability of CI Tau
results from reflex motion induced by a $\sim$11$-$12~M$_{Jup}$ planet.
Furthermore, we find that the flux in the $+$200~km~s$^{-1}$ region of 
CI Tau's H$\alpha$ emission line varies with an apparent periodicity of 
$\sim$9 days, suggestive of detection of accretion onto the planet 
at a particular orbital phase.  While the period of $\sim 9$ days is
strongly detected in our data, the large level of astrophysical noise means
that some of the orbital parameters (e.g., the eccentricity) are not well
determined.  A firmer detection of the rotation period of the star is also
needed.  As a result, more observations of CI Tau are critical to confirm this important result,
as well as additional observations to detect giant planets around other young stars.
An excess in the {\it young} hot Jupiter population may indicate the prevalence of
destructive mechanisms which result in the relative paucity of massive, short-period
planets around main-sequence stars, $\sim$1\% (e.g., Wright et al. 2012).
A massive planet in a $\sim$9 period day orbit around a 2 Myr old star places
strong constraints on planet formation and migration time scales.  It is key to our
understanding of exoplanet evolution to determine how common such systems are.

We thank the IRTF TOs Dave Griep, Bill Golisch, and Eric Volquardsen and SAs
John Rayner, Mike Connelly, and Bobby Bus,
the Keck Observatory OAs Cynthia Wilburn and Heather Hershley and SAs 
Scott Dahm and Greg Wirth, 
KPNO staff Dave Summers, Di Harmer, and Dick Joyce,
and Dave Doss of McDonald Observatory for their exceptional observing support
over the many years of this program.
LP is grateful to Peter Bodenheimer, Joe Llama, Evgenya Shkolnik, and Ben Zuckerman for insightful discussions.
Partial support for this research was provided by the SIM Young Planets Key Project and
by NASA Origins grants 05-SSO05-86 and 07-SSO07-86 to LP.  Additional
support for this work was provided by the NSF through grant AST-1212122
made to Rice University.  We are grateful to 
the Arizona Space Grant consortium for support of JNM's participation in this work.  We wish to thank an anonymous referee for many helpful comments that
improved the manuscript.
This work made use of the SIMBAD reference database, the NASA
Astrophysics Data System, and the data products from the Two Micron All
Sky Survey, which is a joint project of the University of Massachusetts
and the Infrared Processing and Analysis Center/California Institute
of Technology, funded by the National Aeronautics and Space
Administration and the National Science Foundation.
This work made use of the Immersion Grating Infrared Spectrograph (IGRINS) that
was developed under a collaboration between the University of Texas at Austin
and the Korea Astronomy and Space Science Institute (KASI) with the financial
support of the US National Science Foundation under grant AST-1229522, of the
University of Texas at Austin, and of the Korean GMT Project of KASI.
Some data presented herein were obtained at the W. M. Keck
Observatory, which is operated as a scientific partnership among the California Institute of Technology,
the University of California, and the National Aeronautics and Space Administration. The Observatory
was made possible by the generous financial support of the W. M. Keck Foundation.
The authors recognize and acknowledge the
significant cultural role that the summit of Mauna Kea
plays within the indigenous Hawaiian community.  We are
grateful for the opportunity to conduct observations from this special mountain.

\clearpage

\begin{deluxetable}{lcccc} 
\tablecaption{CI Tau Infrared Spectroscopy}
%\rotate
%\tabletypesize{\footnotesize}
\tablewidth{0pt}
\tablehead{
\colhead{Julian} & \colhead{RV} & \colhead{$\sigma_{RV}$} & & \\
	\colhead{Date} & \colhead{(km s$^{-1}$)}  & \colhead{(km s$^{-1}$)} & \colhead{$r_K$} & \colhead{$\sigma_{r_K}$}
}
\startdata
CSHELL\\
\hline
2455156.098 & -0.02 & 0.21 & 2.33 & 0.17 \\
2455158.116 & -0.33 & 0.20 & 1.84 & 0.12 \\
2455160.107 & -0.62 & 0.14 & 2.46 & 0.11 \\
2455235.900 & -0.49 & 0.24 & 1.94 & 0.16 \\
2455236.851 & -0.07 & 0.16 & 2.37 & 0.12 \\
2455237.879 &  0.40 & 0.14 & 1.90 & 0.08 \\
2455238.884 &  1.06 & 0.14 & 1.64 & 0.07 \\
2455239.892 &  1.26 & 0.12 & 1.57 & 0.05 \\
2455240.889 &  0.27 & 0.23 & 1.49 & 0.11 \\
2455241.866 &  0.70 & 0.14 & 1.46 & 0.06 \\
2455242.863 &  0.52 & 0.13 & 1.56 & 0.05 \\
2456258.027 &  0.45 & 0.16 & 1.55 & 0.07 \\
2456258.945 & -1.02 & 0.18 & 1.62 & 0.09 \\
2456259.117 & -1.40 & 0.22 & 2.30 & 0.17 \\
2456259.840 & -0.48 & 0.11 & 1.38 & 0.04 \\
2456260.879 & -0.76 & 0.14 & 1.49 & 0.06 \\
2456263.769 &  0.28 & 0.13 & 1.19 & 0.04 \\
2456622.774 & -0.61 & 0.08 & 0.00 & 0.01 \\
2456623.785 &  0.14 & 0.10 & 0.13 & 0.01 \\
2456624.859 &  1.01 & 0.11 & 0.77 & 0.03 \\
2456625.881 &  0.88 & 0.19 & 1.11 & 0.02 \\
2456626.911 &  0.16 & 0.40 & 0.97 & 0.10 \\
2456690.895 & -0.29 & 0.31 & 0.35 & 0.02 \\
2456693.750 & -0.45 & 0.63 & 1.51 & 0.40 \\
2456696.769 &  0.74 & 0.15 & 0.95 & 0.08 \\
2456697.773 &  0.20 & 0.17 & 0.94 & 0.05 \\
2456698.747 &  0.13 & 0.46 & 0.90 & 0.14 \\
2456699.748 &  0.08 & 0.14 & 0.93 & 0.05 \\
2456701.745 & -0.52 & 0.40 & 1.05 & 0.06 \\
2456714.739 &  0.67 & 0.10 & 0.94 & 0.01 \\
2456715.737 &  1.98 & 0.22 & 0.98 & 0.03 \\
2456716.743 & -0.03 & 0.23 & 0.98 & 0.07 \\
2456717.741 &  1.22 & 0.34 & 0.86 & 0.06 \\
2456723.740 &  0.75 & 0.33 & 0.51 & 0.01 \\
 \hline\\
 NIRSPEC\\
\hline
2455251.727 &  -0.14 &  0.07 & 1.49 & 0.03 \\
2455255.730 &  -0.86 &  0.05 & 1.64 & 0.03 \\
2455522.880 &  -0.26 &  0.05 & 1.73 & 0.03 \\
\hline\\
Phoenix\\
\hline
2456350.633\tablenotemark{a} & -1.24 & 0.06 & 0.34 & 0.01 \\
2456351.614\tablenotemark{a} & -1.29 & 0.07 & 0.41 & 0.01 \\
2456352.610\tablenotemark{a} & -0.42 & 0.06 & 0.40 & 0.01 \\
2456353.607\tablenotemark{a}  & 0.27 & 0.07 & 0.45 & 0.01 \\
2456354.668\tablenotemark{a}  & 0.27 & 0.07 & 0.22 & 0.01 \\
2456605.732 & -0.56 & 0.06 & 0.13 & 0.01 \\
2456606.810  & 0.48 & 0.06 & 0.33 & 0.01 \\
2456607.778   & 1.40 & 0.07 & 0.29 & 0.01 \\
2456608.761  & 0.88 & 0.07 & 0.26 & 0.01 \\
2456610.769 & -1.41 & 0.06 & 0.26 & 0.01 \\
2456643.930  &  1.32 & 0.07 & 0.00 & 0.01 \\
2456644.875 & -1.18 & 0.13 & 0.08 & 0.01 \\
2456645.771  & 0.31 & 0.08 & 0.41 & 0.01 \\
2456661.857   & 2.60 & 0.16 & 0.53 & 0.01 \\
2456662.840 & -2.01 & 0.13 & 0.48 & 0.01 \\
2456663.817 & -0.73 & 0.08 & 0.29 & 0.01 \\
2456664.723 & -1.01 & 0.07 & 0.48 & 0.01 \\
2456666.708 & -0.79 & 0.07 & 0.41 & 0.01 \\
2456667.791 & -0.49 & 0.07 & 0.24 & 0.01 \\
\hline\\
IGRINS\\
\hline
2456925.895 & -0.89 & 0.32 & 1.45 & 0.01 \\
2456940.827 &  2.14 & 0.32 & 1.18 & 0.04 \\
2456984.776 &  0.90 & 0.34 & 1.29 & 0.07 \\
2456985.924 & -0.04 & 0.15 & 0.88 & 0.03 \\
2456986.908 &  0.95 & 0.20 & 0.87 & 0.04 \\
2456987.851 & -1.03 & 0.31 & 1.08 & 0.05 \\
2456988.855 &  0.36 & 0.13 & 1.33 & 0.03 \\
2456989.788 & -1.07 & 0.13 & 1.56 & 0.01 \\
2456990.813 & -1.19 & 0.27 & 1.45 & 0.07 \\
2456991.704 & -1.42 & 0.23 & 1.39 & 0.07 \\
2456992.686 &  0.87 & 0.24 & 1.19 & 0.04 \\
2456993.745 &  0.20 & 0.35 & 1.38 & 0.05 \\
2456996.826 &  0.43 & 0.21 & 1.22 & 0.06 \\
2456997.686 & -0.39 & 0.28 & 1.70 & 0.10 \\
2457029.577 & -0.77 & 0.54 & 1.03 & 0.07 \\
\enddata

\tablenotetext{a}{For these nights, Phoenix was mounted on the KPNO 4-meter; all other Phoenix data were taken with the KPNO 2.1-meter.}

\end{deluxetable}

\clearpage

\begin{deluxetable}{lrrrrrrrr} 
\tablecaption{CI Tau Optical Spectroscopy}
%\rotate
%\tabletypesize{\footnotesize}
\tablewidth{0pt}
\tablehead{
   \colhead{Julian} & 
   \colhead{RV} & 
   \colhead{$\sigma_{RV}$} &
   \colhead{\ion{Ca}{2} RV} & 
	\colhead{$\sigma_{Ca {\sc II}}$} &
   \colhead{\ion{He}{1} RV} & 
	\colhead{$\sigma_{He {\sc I}}$} &
   \colhead{} & 
   \colhead{}\\[0.2ex]
   \colhead{Date} & 
   \colhead{(km s$^{-1}$)} & 
   \colhead{(km s$^{-1}$)} &
   \colhead{(km s$^{-1}$)} &
   \colhead{(km s$^{-1}$)} &
   \colhead{(km s$^{-1}$)} &
   \colhead{(km s$^{-1}$)} &
   \colhead{$r$} &
   \colhead{$\sigma_{r}$}
}
\startdata
2453367.805 & -0.46 & 0.22 & 19.21 & 0.46 & 22.79 & 0.93 & 0.09 &0.01\\
2453696.899 & -0.03 & 0.26 & 17.66 & 0.24 & 21.82 & 0.60 & 0.12 & 0.01\\
2453770.777 & 1.05 & 0.25 & 18.43 & 0.40 & 25.19 & 0.94 & 0.66 & 0.02\\
2454141.782 & \nodata & \nodata & 16.72 & 0.40 & 25.38 & 0.70 & \nodata & \nodata\\
2454424.950 & 0.12 & 0.23 & 17.49 & 0.37 & 15.85 & 1.22 & 0.54 & 0.01\\
2455159.848 & -0.71 & 0.21 & 17.89 & 0.60 & 19.47 & 1.51 & 0.31 & 0.02\\
2455160.819 & 0.59 & 0.14 & 17.53 & 0.53 & 19.67 & 0.61 & 0.79 & 0.02\\
2455161.787 & 0.96 & 0.20 & 15.44\tablenotemark{a} & 0.31 & 20.26 & 0.64 & 0.94 & 0.02\\
2455162.834 & -0.72 & 0.22 & 17.74 & 0.41 & 21.06 & 1.10 & 1.02 & 0.03\\
2455163.908 & -1.32 & 0.28 & 18.84 & 0.35 & 22.10 & 1.00 & 0.75 & 0.02\\
2455164.998 & -1.75 & 0.26 & 15.43\tablenotemark{b} & 0.55 & 23.43 & 3.05 & 0.57 & 0.03\\
2456251.697 & -0.44 & 0.28 & 18.44 & 0.34 & 21.42 & 0.72 & 0.53 & 0.02\\
2456252.678 & -1.24 & 0.38 & 16.00 & 0.48 & 22.24 & 1.01 & 0.33 & 0.02\\
2456253.844 & -0.68 & 0.24 & 17.23 & 0.51 & 19.79 & 1.08 & 0.27 & 0.01\\
2456254.680 & -0.10 & 0.17 & 17.90 & 0.20 & 21.14 & 0.81 & 0.07 & 0.01\\
2456255.703 & -0.06 & 0.18 & 17.65 & 0.39 & 22.85 & 0.80 & 0.10 & 0.01\\
2456256.669 & -0.40 & 0.18 & 17.61 & 0.38 & 26.37 & 0.99 & 0.51 & 0.01\\
2456257.669 & \nodata & \nodata & 18.57 & 0.23 & 25.68 & 2.17 & \nodata & \nodata\\
2456257.905 & \nodata & \nodata & 19.88 & 0.55 & 18.70 & 1.62 & \nodata & \nodata\\
2456605.692 & 0.22 & 0.35 & 18.35 & 0.26 & 20.12 & 0.66 & 0.03 & 0.01\\
2456605.906 & -0.06 & 0.35 & 19.48 & 0.29 & 21.90 & 0.63 & 0.00\tablenotemark{c} & 0.00\\
2456606.680 & 0.76 & 0.40 & 17.00 & 0.33 & 20.45 & 0.57 & 0.17 & 0.01\\
2456606.927 & 0.45 & 0.34 & 18.54 & 0.31 & 21.73 & 0.47 & 0.18 & 0.01\\
2456607.678 & 1.86 & 0.50 & 16.88 & 0.25 & 21.81 & 0.52 & 0.59 & 0.02\\
2456607.871 & 1.04 & 0.55 & 18.43 & 0.33 & 22.54 & 0.48 & 0.37 & 0.01\\
2456608.694 &  1.15 & 0.66 & 15.46 & 0.98 & 18.41 & 1.99 & 0.69 & 0.04\\
2456608.743 & 0.63 & 0.52 & 17.92 & 0.30 & 22.22 & 0.64 & 0.68 & 0.02\\
2456608.949 & 0.61 & 0.53 & 17.46\tablenotemark{b} & 0.32 & 21.88 & 0.66 & 0.60 & 0.02\\
2456611.659 & -1.54 & 0.56 & 16.86 & 0.32 & 20.09 & 1.04 & 0.19 & 0.01\\
\enddata
\tablenotetext{a}{Strong inverse P-Cygni absorption affecting \ion{Ca}{2} 
profile.  The \ion{Ca}{2} RV likely biased to a lower value.}
\tablenotetext{b}{Weaker inverse P-Cygni absorption affecting \ion{Ca}{2} 
profile.  The \ion{Ca}{2} RV is possibly biased to a lower value.}
\tablenotetext{c}{Spectrum used as reference for relative veiling measurements.
Veling is 0.0 by definition.}
\end{deluxetable}

\clearpage

\begin{deluxetable}{lll} 
\tablecaption{CI Tau Photometry}
%\rotate
%\tabletypesize{\footnotesize}
\tablewidth{0pt}
\tablehead{
\colhead{Heliocentric} & \colhead{V} & \colhead{$\sigma_V$}\\
\colhead{Julian Date} & \colhead{(mag)}  & \colhead{(mag)}
}
\startdata
2456238.73400     & 12.883    & 0.004\\
2456238.73618     & 12.886    & 0.004\\
2456238.73836     & 12.885    & 0.004\\
2456238.87180     & 12.919    & 0.004\\
2456238.87398     & 12.921    & 0.004\\
2456239.01481     & 12.919    & 0.004\\
2456239.01699     & 12.917    & 0.004\\
2456239.01917     & 12.919    & 0.004\\
2456239.02135     & 12.917    & 0.004\\
2456239.02353     & 12.918    & 0.004\\
2456239.02571     & 12.907    & 0.004\\
2456239.71033     & 12.971    & 0.004\\
2456239.71251     & 12.977    & 0.004\\
2456239.71468     & 12.975    & 0.004\\
2456239.84985     & 12.984    & 0.004\\
2456239.85203     & 12.978    & 0.004\\
2456239.85421     & 12.983    & 0.004\\
2456240.02371     & 12.978     & 0.005\\
2456240.02588     & 12.969     & 0.005\\
2456240.02807     & 12.961     & 0.005\\
2456240.04286     & 12.979      & 0.007\\
2456240.04504     & 12.976      & 0.008\\
2456240.04722     & 12.975      & 0.009\\
2456251.63990     & 12.863     & 0.005\\
2456251.64208     & 12.865     & 0.005\\
2456251.72535     & 12.824    & 0.004\\
2456251.72754     & 12.822    & 0.004\\
2456251.81324     & 12.938    & 0.004\\
2456251.81541     & 12.943    & 0.004\\
2456251.90858     & 12.964    & 0.004\\
2456251.91076     & 12.966    & 0.004\\
2456251.97975     & 12.969    & 0.004\\
2456251.98193     & 12.973    & 0.004\\
2456252.01878     & 12.966    & 0.004\\
2456252.02095     & 12.967    & 0.004\\
2456252.04931     & 12.965      & 0.007\\
2456252.05150     & 12.978      & 0.008\\
2456252.63705     & 12.992     & 0.005\\
2456252.63924     & 12.999     & 0.005\\
2456252.72251     & 12.903    & 0.004\\
2456252.72468     & 12.894    & 0.004\\
2456252.81035     & 12.841    & 0.004\\
2456252.81253     & 12.850    & 0.004\\
2456252.90557     & 12.934    & 0.004\\
2456252.90774     & 12.943    & 0.004\\
2456252.97758     & 12.934    & 0.004\\
2456252.97975     & 12.935    & 0.004\\
2456253.01663     & 12.941    & 0.004\\
2456253.01881     & 12.957    & 0.004\\
2456253.04729     & 12.945     & 0.005\\
2456253.04947     & 12.940     & 0.006\\
2456253.64016     & 12.908     & 0.006\\
2456253.64234     & 12.889     & 0.006\\
2456253.72515     & 12.939     & 0.005\\
2456253.72734     & 12.941     & 0.005\\
2456253.81376     & 12.963    & 0.004\\
2456253.81594     & 12.962    & 0.004\\
2456253.90444     & 13.031    & 0.004\\
2456253.90662     & 13.031    & 0.004\\
2456253.97342     & 13.056     & 0.006\\
2456254.01777     & 13.064    & 0.004\\
2456254.01995     & 13.060    & 0.004\\
2456254.04838     & 13.081     & 0.006\\
2456254.05056     & 13.068      & 0.007\\
2456254.63743     & 13.172     & 0.006\\
2456254.63961     & 13.172     & 0.006\\
2456254.72235     & 13.201     & 0.005\\
2456254.72453     & 13.205     & 0.005\\
2456254.81096     & 13.219     & 0.005\\
2456254.81314     & 13.217     & 0.005\\
2456254.90172     & 13.235    & 0.004\\
2456254.90390     & 13.240    & 0.004\\
2456254.96857     & 13.243    & 0.004\\
2456254.97074     & 13.243    & 0.004\\
2456255.01516     & 13.253    & 0.004\\
2456255.01734     & 13.256    & 0.004\\
2456255.04579     & 13.237     & 0.005\\
2456255.04796     & 13.225     & 0.006\\
2456255.62647     & 13.244      & 0.007\\
2456255.62865     & 13.250     & 0.006\\
2456255.71147     & 13.260     & 0.005\\
2456255.71365     & 13.266     & 0.005\\
2456255.79863     & 13.254     & 0.005\\
2456255.80081     & 13.258     & 0.005\\
2456255.90262     & 13.279     & 0.005\\
2456255.90481     & 13.280     & 0.005\\
2456255.96618     & 13.279    & 0.004\\
2456255.96836     & 13.281    & 0.004\\
2456256.00508     & 13.291    & 0.004\\
2456256.00725     & 13.286    & 0.004\\
2456256.04240     & 13.284     & 0.005\\
2456256.04458     & 13.287     & 0.005\\
2456256.64926     & 13.070      & 0.008\\
2456256.65144     & 13.062      & 0.008\\
2456256.75939     & 13.077     & 0.006\\
2456256.76157     & 13.074     & 0.006\\
2456256.87745     & 13.011     & 0.006\\
2456256.87963     & 13.008     & 0.006\\
2456256.96075     & 13.031     & 0.006\\
2456256.96293     & 13.030     & 0.006\\
2456257.03305     & 13.028     & 0.005\\
2456257.03523     & 13.041     & 0.005\\
2456267.62628     & 12.954      & 0.007\\
2456267.62846     & 12.970      & 0.007\\
2456267.69684     & 12.956     & 0.005\\
2456267.69902     & 12.962     & 0.005\\
2456267.78586     & 13.035     & 0.006\\
2456267.78804     & 13.041     & 0.006\\
2456267.87890     & 13.058     & 0.005\\
2456267.88108     & 13.056    & 0.004\\
2456267.94838     & 13.024     & 0.005\\
2456267.95056     & 13.022     & 0.005\\
2456268.01312     & 13.001     & 0.006\\
2456268.01530     & 13.002     & 0.006\\
2456268.61656     & 12.995    & 0.004\\
2456268.61874     & 12.989    & 0.004\\
2456268.67150     & 12.994    & 0.004\\
2456268.67368     & 12.995    & 0.004\\
2456268.74206     & 13.007    & 0.004\\
2456268.74425     & 13.009     & 0.004\\
2456268.81746     & 13.046     & 0.004\\
2456268.81964     & 13.052     & 0.004\\
2456268.88553     & 13.083     & 0.004\\
2456268.88771     & 13.090     & 0.004\\
2456268.95393     & 13.065     & 0.005\\
2456268.95611     & 13.060     & 0.005\\
2456269.01889     & 13.036     & 0.006\\
2456269.02107     & 13.021     & 0.006\\
2456269.58786     & 13.040     & 0.004\\
2456269.59004     & 13.036     & 0.004\\
2456269.66539     & 13.045     & 0.004\\
2456269.66757     & 13.049     & 0.004\\
2456269.74140     & 13.004     & 0.004\\
2456269.74358     & 13.000     & 0.004\\
2456269.82240     & 13.024     & 0.004\\
2456269.82457     & 13.018     & 0.004\\
2456269.85479     & 13.019     & 0.004\\
2456269.85697     & 13.016     & 0.004\\
2456269.93296     & 13.012     & 0.004\\
2456269.93514     & 13.014     & 0.004\\
2456269.99843     & 13.019     & 0.005\\
2456270.00061     & 13.013     & 0.005\\
2456270.58799     & 13.084     & 0.005\\
2456270.59017     & 13.093     & 0.005\\
2456270.65912     & 13.112     & 0.004\\
2456270.66130     & 13.113     & 0.004\\
2456270.72622     & 13.098     & 0.004\\
2456270.72840     & 13.100     & 0.004\\
2456270.79856     & 13.080     & 0.004\\
2456270.80074     & 13.084     & 0.004\\
2456270.82714     & 13.092     & 0.004\\
2456270.82932     & 13.099     & 0.004\\
2456270.89863     & 13.100     & 0.004\\
2456270.90082     & 13.097     & 0.004\\
2456270.96193     & 13.096     & 0.005\\
2456271.58812     & 13.131     & 0.005\\
2456271.59029     & 13.120     & 0.005\\
2456271.65918     & 13.100     & 0.005\\
2456271.66136     & 13.105     & 0.005\\
2456271.72620     & 13.110     & 0.004\\
2456271.72838     & 13.111     & 0.004\\
2456271.80053     & 13.079     & 0.004\\
2456271.80271     & 13.079     & 0.004\\
2456271.83071     & 13.108     & 0.004\\
2456271.83289     & 13.105     & 0.004\\
2456271.89573     & 13.090     & 0.004\\
2456271.89791     & 13.089     & 0.004\\
2456271.95901     & 13.106     & 0.005\\
2456271.96119     & 13.096     & 0.005\\
2456272.58097     & 12.915     & 0.005\\
2456272.58315     & 12.911     & 0.004\\
2456272.63205     & 12.934     & 0.004\\
2456272.63423     & 12.929     & 0.004\\
2456272.68054     & 12.883     & 0.004\\
2456272.68272     & 12.880     & 0.004\\
2456272.74247     & 12.865     & 0.004\\
2456272.74465     & 12.870     & 0.004\\
2456272.79122     & 12.904     & 0.004\\
2456272.79340     & 12.911     & 0.004\\
2456272.83942     & 12.932     & 0.004\\
2456272.84160     & 12.924     & 0.004\\
2456272.84863     & 12.923     & 0.004\\
2456272.85081     & 12.931     & 0.004\\
2456272.89251     & 13.003     & 0.004\\
2456272.89469     & 12.993     & 0.004\\
2456272.93692     & 12.984     & 0.004\\
2456272.93910     & 12.991     & 0.004\\
2456272.98148     & 12.991     & 0.004\\
2456272.98366     & 12.996     & 0.004\\
\enddata

\end{deluxetable}

\clearpage

%\clearpage

\begin{deluxetable}{lcc}
\tablewidth{0pt}
\tablecaption{CI Tau Orbital Properties and Inferred Mass of CI Tau b
	\label{tbl:orbitalelements}}
\tablehead{
   \colhead{ } &
   \colhead{Using IR} &
   \colhead{Using IR \&}\\[0.2ex]
   \colhead{Parameter} & 
   \colhead{RVs} &
   \colhead{Optical RVs}
} 
\startdata
P (days) &  8.9965 $\pm$ 0.0327 & 8.9891 $\pm$ 0.0202 \\
K (km s$^{-1}$) & 1.084 $\pm$ 0.250 & 0.950 $\pm$ 0.207 \\
$e$ & 0.40 $\pm$ 0.16 & 0.28 $\pm$ 0.16 \\
$M_p\sin i$ (M$_{Jup}$) & 8.81 $\pm$ 1.71 & 8.08 $\pm$ 1.53  \\
Fit RMS (km s$^{-1}$) & 0.694 & 0.728 \\
\enddata

\end{deluxetable}

\clearpage

\begin{figure*}
\includegraphics[angle=90,width=6.5in]{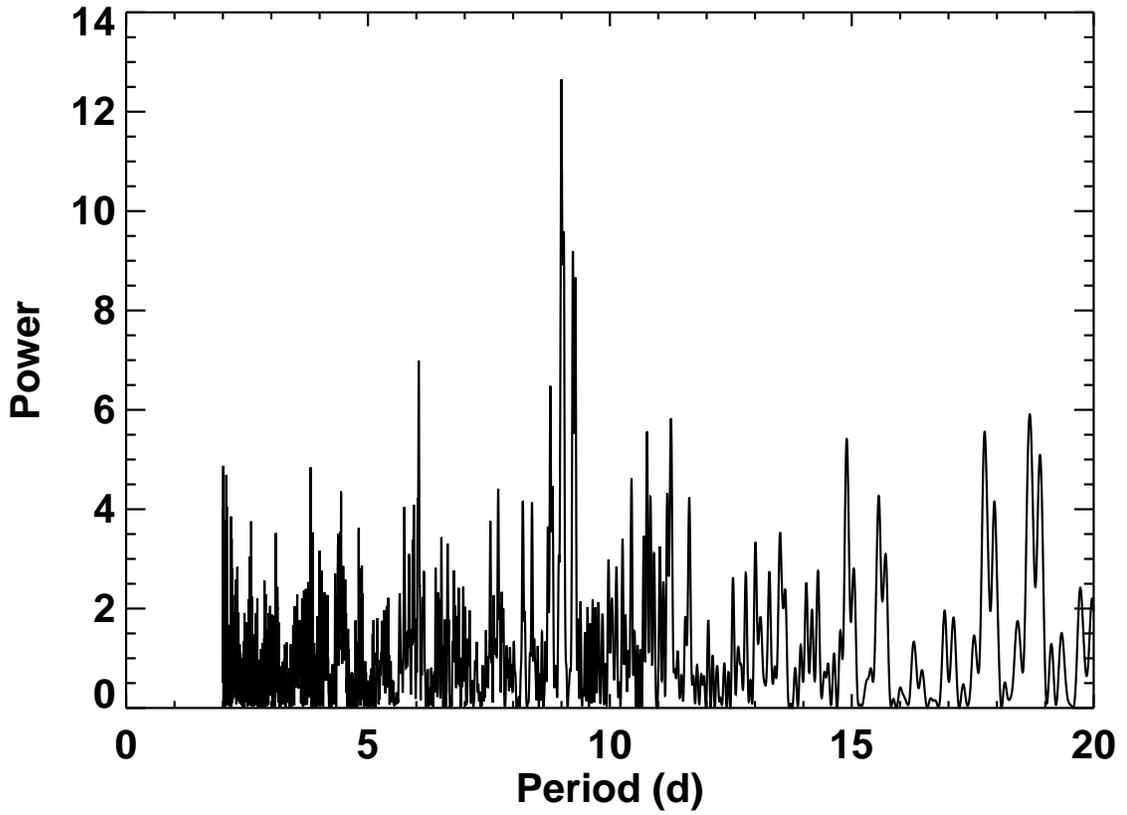}
\caption{Power spectrum for the CI Tau IR spectroscopy based on multiple observations at different telescopes between 2009 November and 2014 November.
{\bf The strongest} peak appears with a 9 day period; the false alarm probability,
calculated for these irregularly sampled data with a Monte Carlo simulation (see text), is 0.001.}
\label{fig1}
\end{figure*}

\clearpage

\begin{figure*}
\includegraphics[angle=90,width=6.5in]{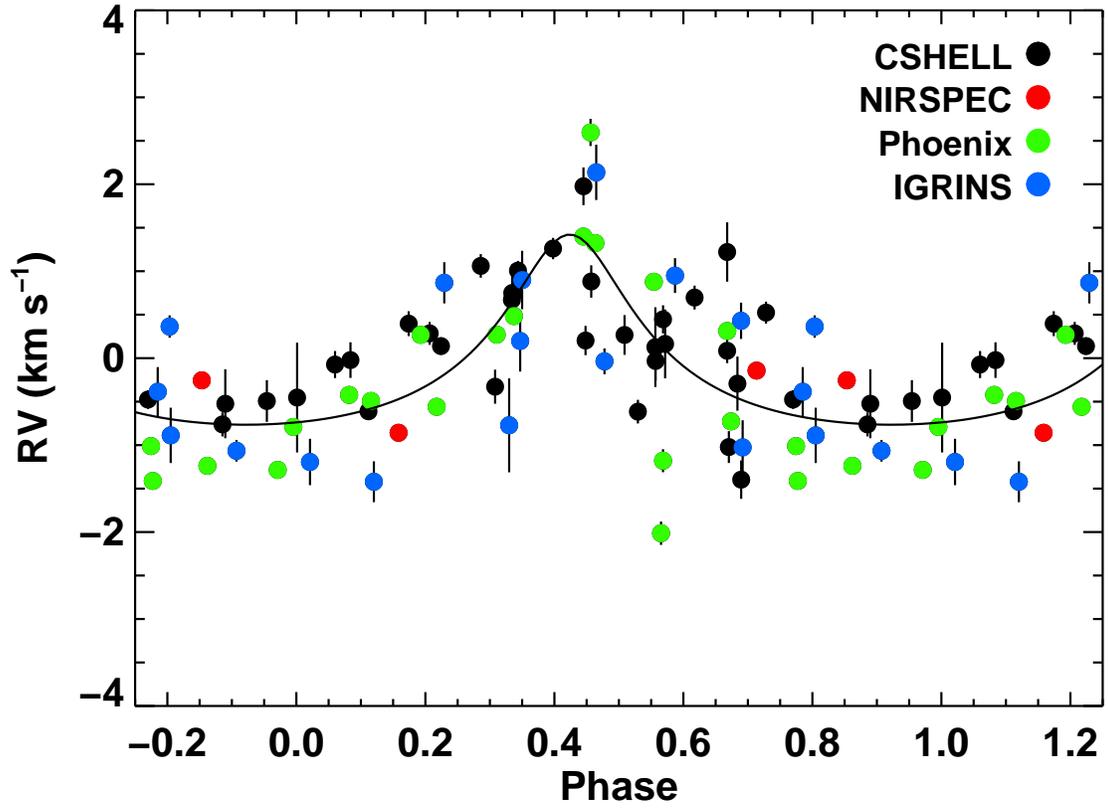}
\caption{RVs for CI Tau based on all IR spectroscopy and phased to a period of
8.9965 days.  The average RV has been subtracted from the data.  Points are 
color coded to indicate the instrument used in the observations (Table 1)}
\label{fig2}
\end{figure*}

\clearpage

\begin{figure*}
\includegraphics[angle=0,width=6.5in]{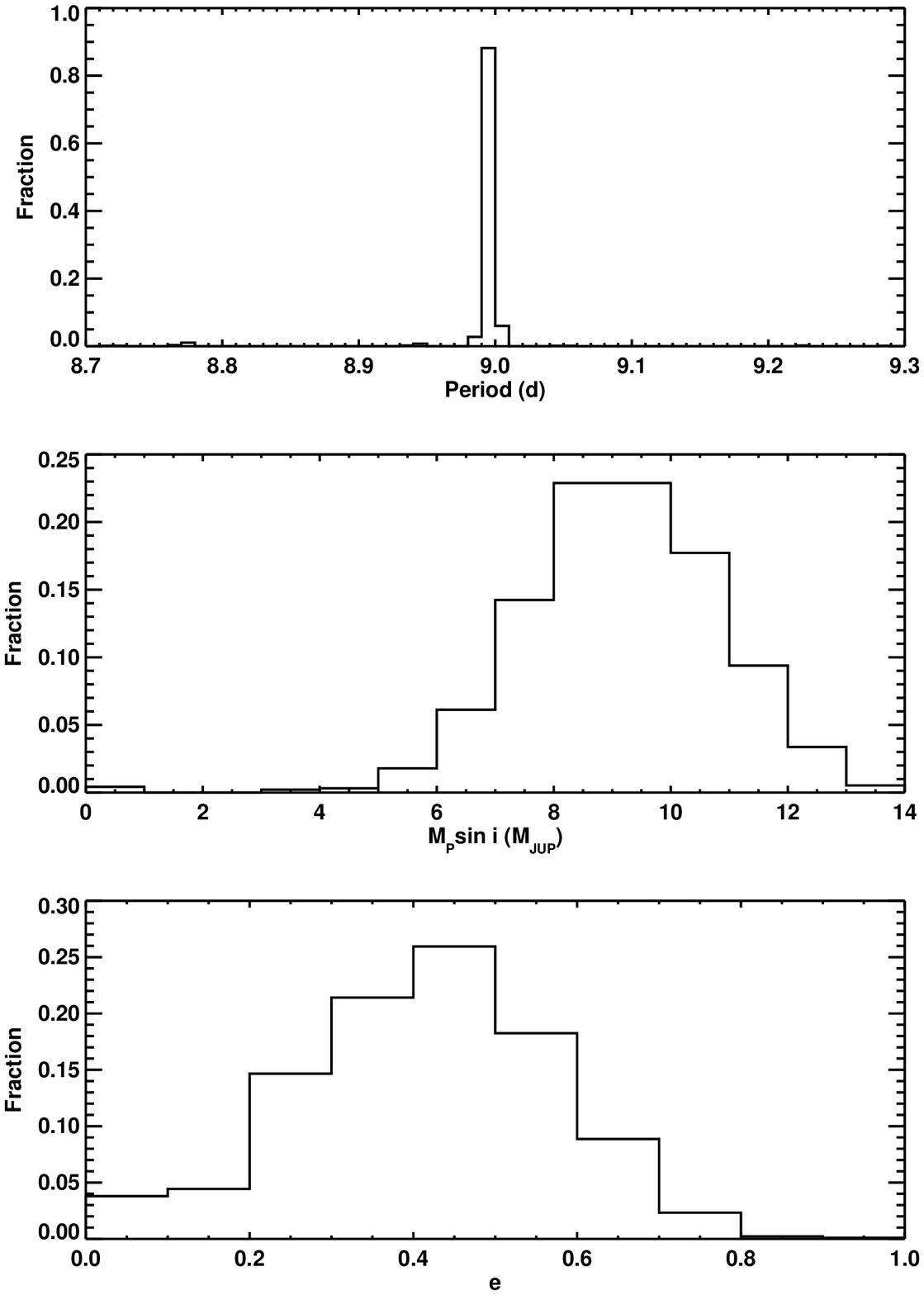}
\caption{The distribution of recovered orbital properties based on the
Monte Carlo simulation of the IR only RV data.  The top panel gives the
recovered period, the middle panel gives the inferred planetary mass, and
the bottom panel give the orbital eccentricity.}
\label{fig3}
\end{figure*}

\clearpage

\begin{figure*}
\includegraphics[angle=0,width=6.0in]{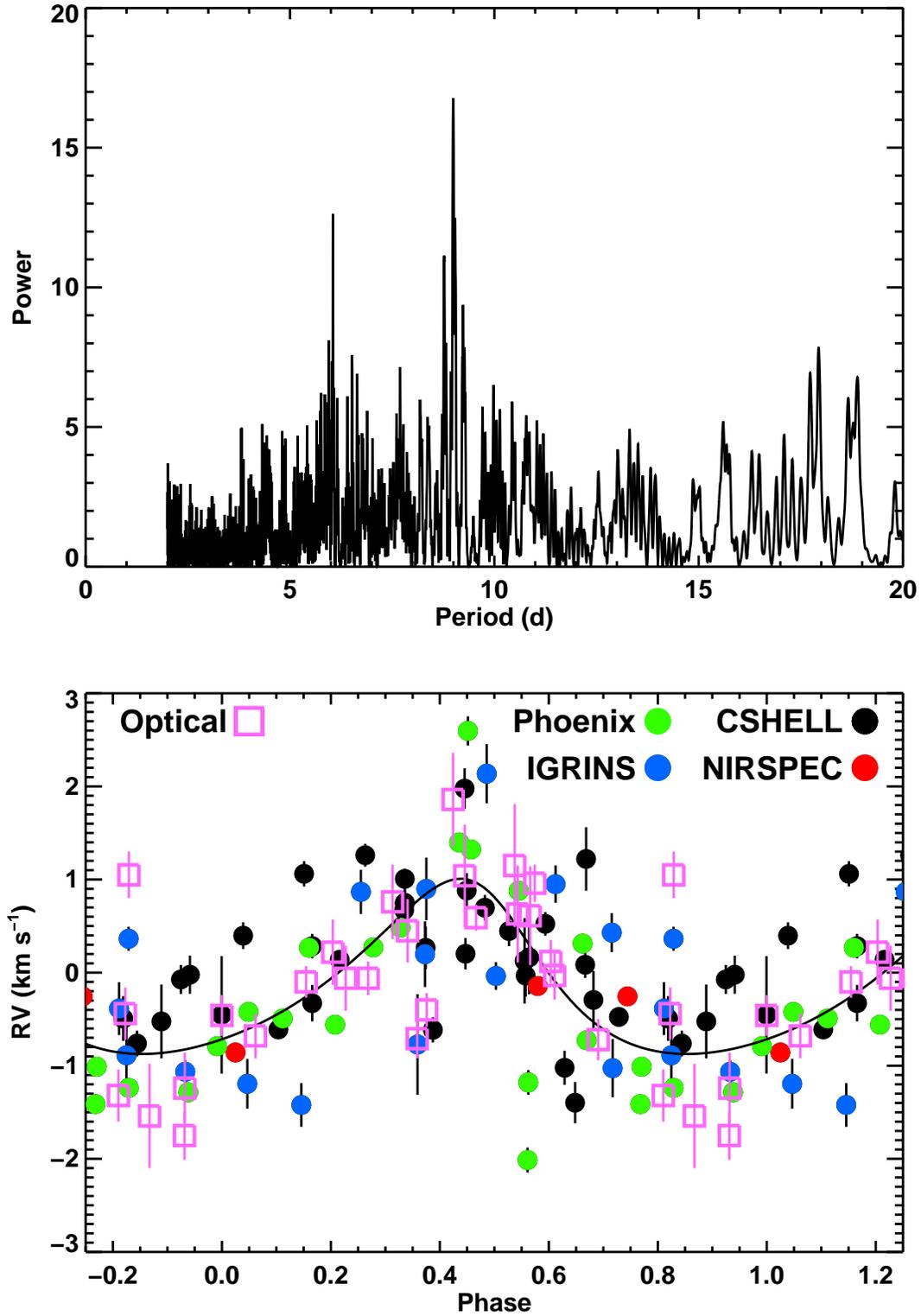}
\caption{The upper panel shows the power spectrum of the combined optical
and IR RV times series.  The peak at $\sim 9$ d has a false alarm
probability of $<10^{-4}$.  The lower panel shows the IR and optical RV 
measurements phased to 8.99 d, determined from the combined RV 
time series.}
\label{fig4}
\end{figure*}

\clearpage

\begin{figure*}
\includegraphics[angle=0,width=5.5in]{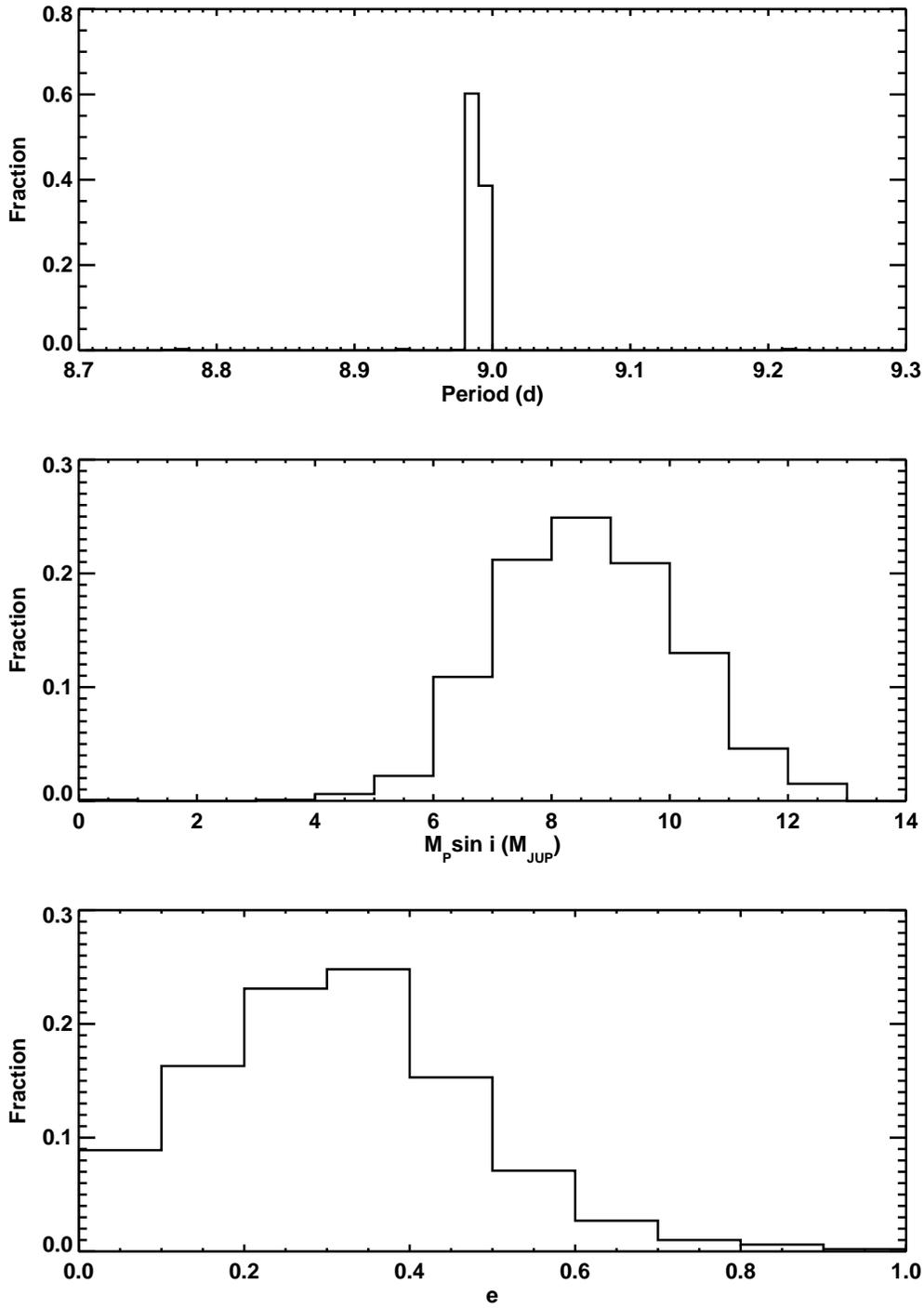}
\caption{The distribution of recovered orbital properties based on the
Monte Carlo simulation of the IR plus optical RV data.  The top panel gives the
recovered period, the middle panel gives the inferred planetary mass, and
the bottom panel give the orbital eccentricity.}
\label{fig5}
\end{figure*}

\clearpage

\begin{figure*}
\includegraphics[angle=90,width=6.5in]{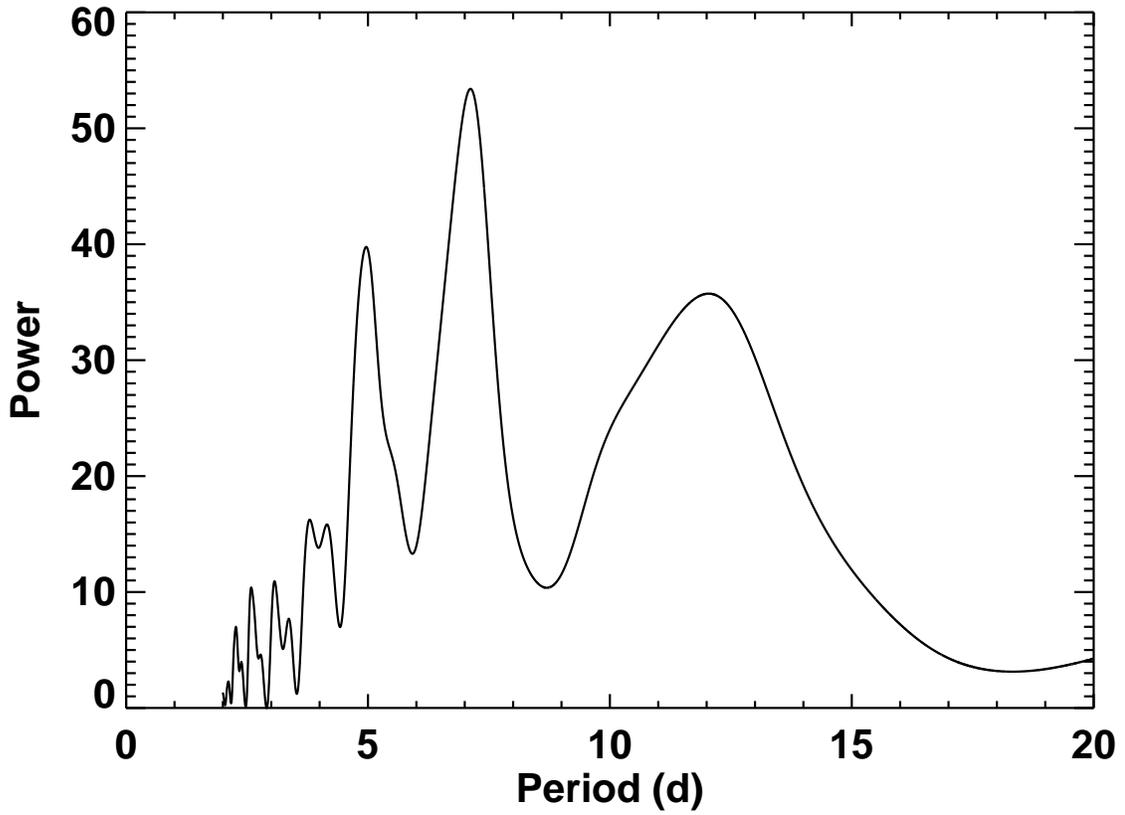}
\caption{Power spectrum for the CI Tau optical photometry based on multiple observations per night (Table 1) over 14 nights in 2012 November and December.
A clear peak appears with a 7.1 day period; the false alarm probability, calculated for these irregularly sampled data with a
Monte Carlo simulation (see text), is $<10^{-4}$.}
\label{fig6}
\end{figure*}

\clearpage

\begin{figure*}
\includegraphics[angle=0,width=5.5in]{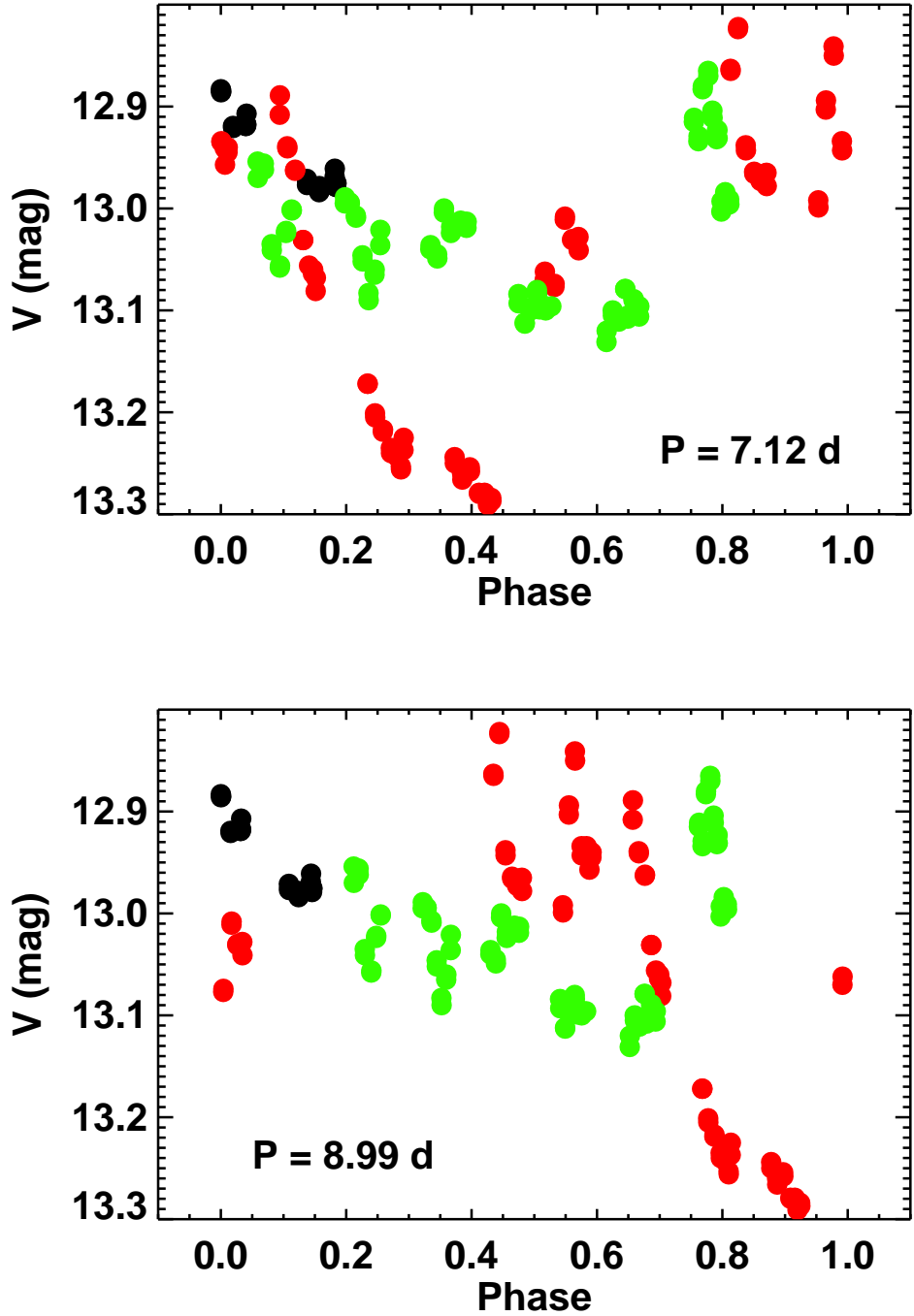}
\caption{The upper panel shows the light curve showing the CI Tau V band 
photometry phased to period of 7.12 days.  Uncertainties are smaller than
the plot symbols.  The black points are from the first observing run, red
are from the second, and green from the third.  The bottom panel shows the
same data, only this time phased to a period of 8.99 days.}
\label{fig7}
\end{figure*}

\clearpage

\begin{figure*}
\includegraphics[angle=90,width=6.5in]{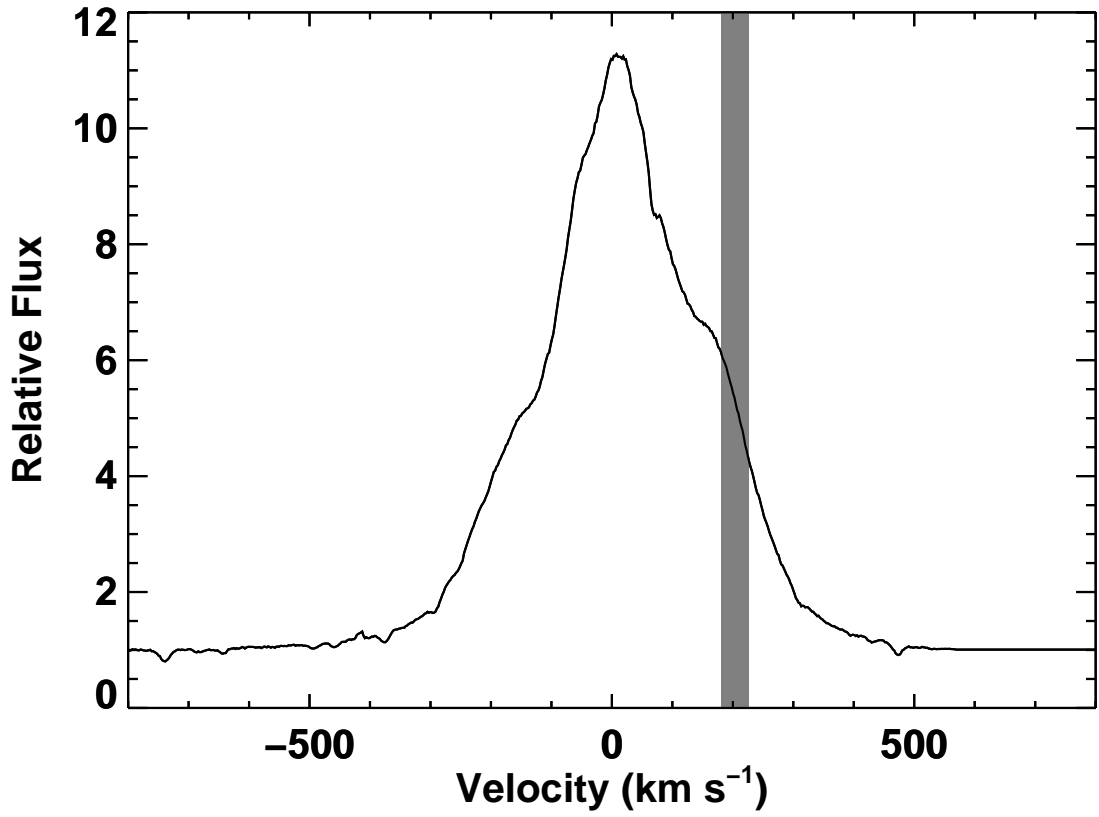}
\caption{The mean H$\alpha$ line profile for CI Tau.  The velocity range
highlighted by the gray bar appears to show significant periodicity with a
period near 9 d (see text).}
\label{fig8}
\end{figure*}

\clearpage

\begin{figure*}
\includegraphics[angle=0,width=5.5in]{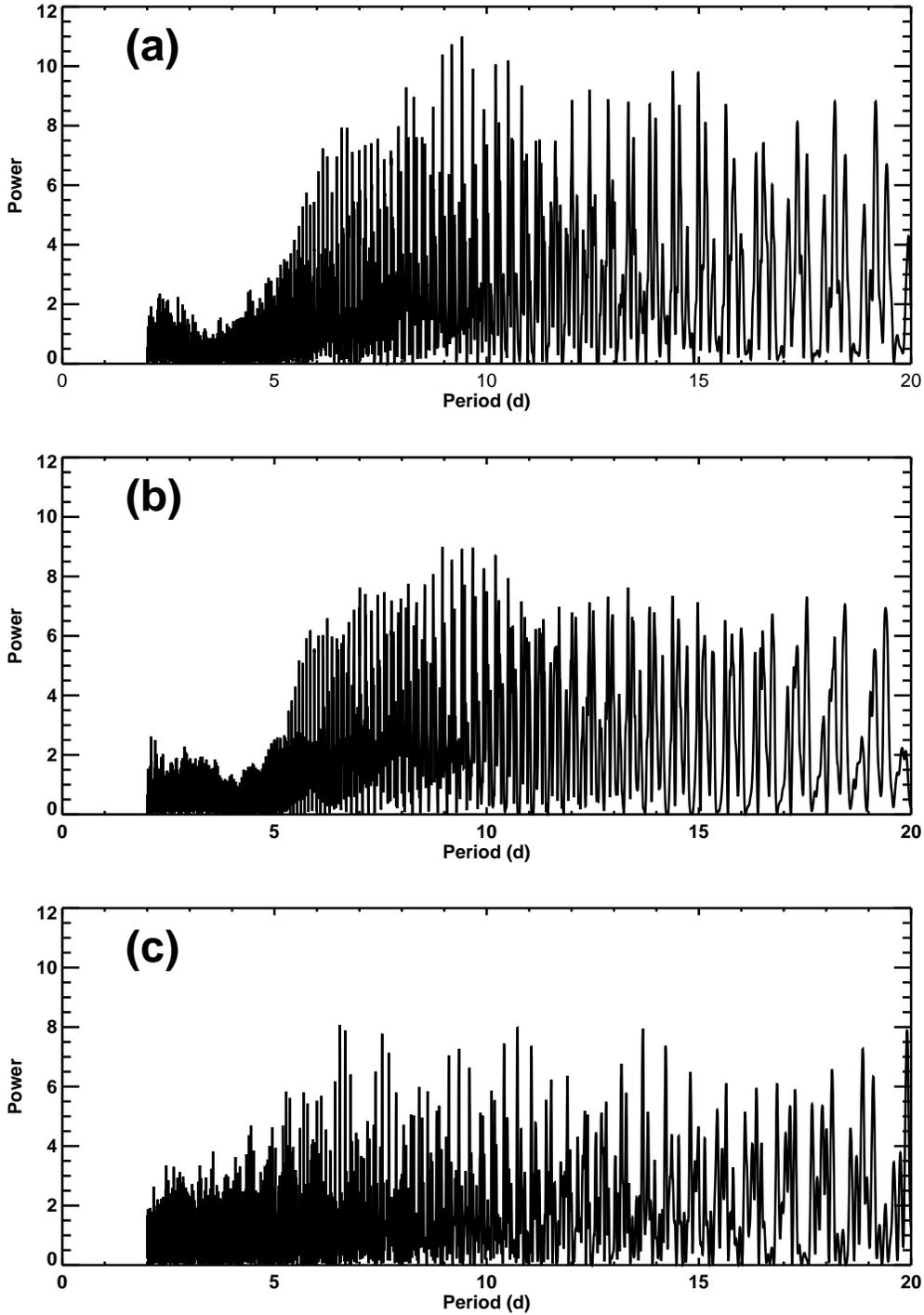}
\caption{(a) The power spectrum of the H$\alpha$ velocity channel around $\sim
200$ km s$^{-1}$ showing the strongest power in the periodogram analysis.
The 3 peaks from $\sim 9.0 - 9.4$ d all have a false alarm probability 
$< 10^{-4}$; however, they likely represent only one actual signal (see text).
(b) and (c) The power spectra of the H$\alpha$ relative flux variations for
the velocity channels around 0 km s$^{-1}$  and $-135$ km s$^{-1}$,
respectively, showing the overall weakening of the power spectrum at other
velocity channels.}
\label{fig9}
\end{figure*}

\clearpage

\begin{figure*}
\includegraphics[angle=0,width=5.5in]{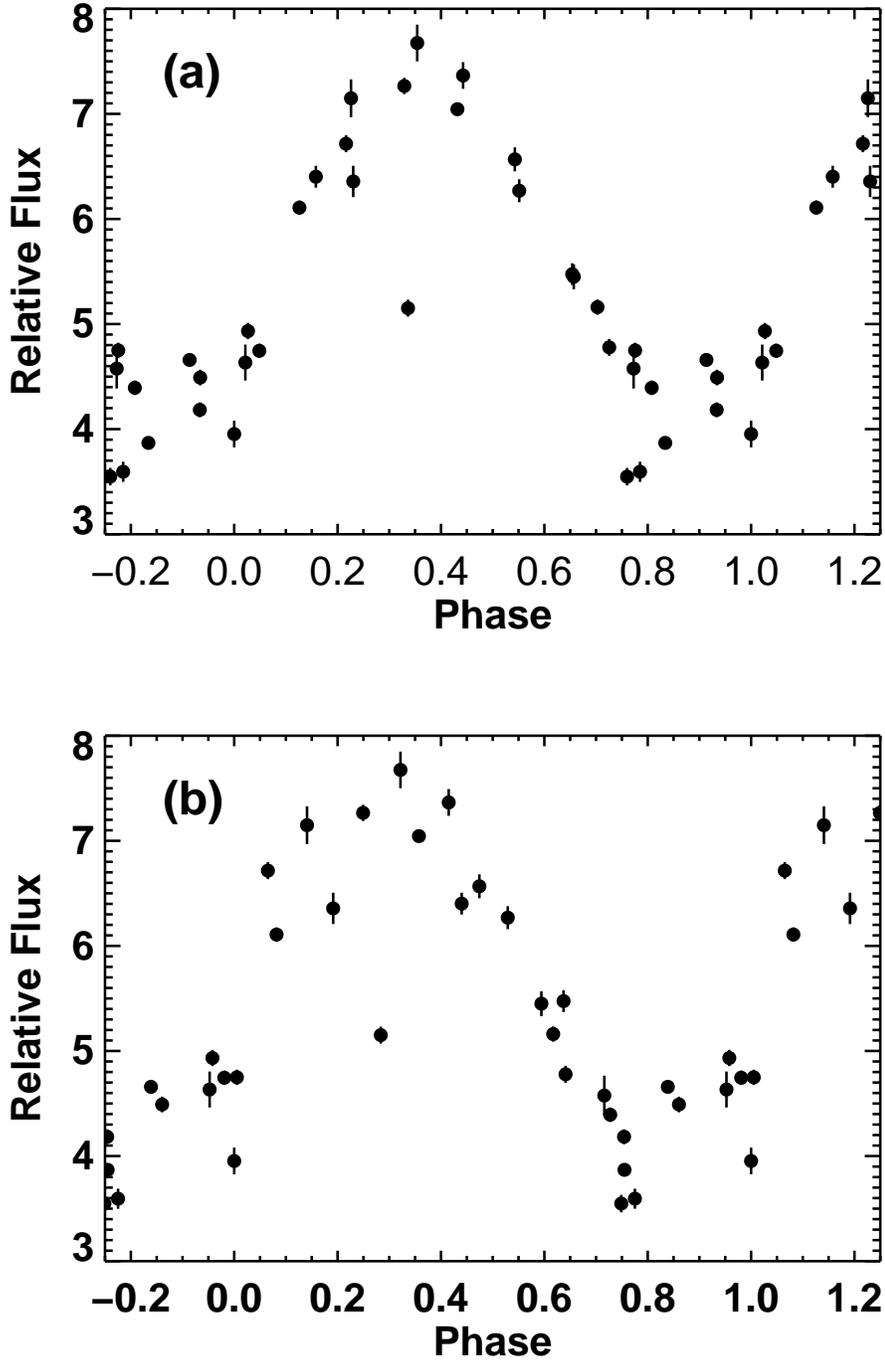}
\caption{(a) The phased ($p = 9.4$ d) H$\alpha$ flux variation curve for the velocity 
channel showing the strongest power, shown in the periodogram in Figure 9a.
(b) The phased H$\alpha$ flux variation curve for the 
same velocity channel shown in Figure 9a but phased to $p = 9.0$ d.  The data phase well at this
period also, indicating that we are not able to narrow down the true period
beyond stating it is likely in the range 9.0 -- 9.4 d.}
\label{fig10}
\end{figure*}

\clearpage

\begin{figure*}
\includegraphics[angle=0,width=5.5in]{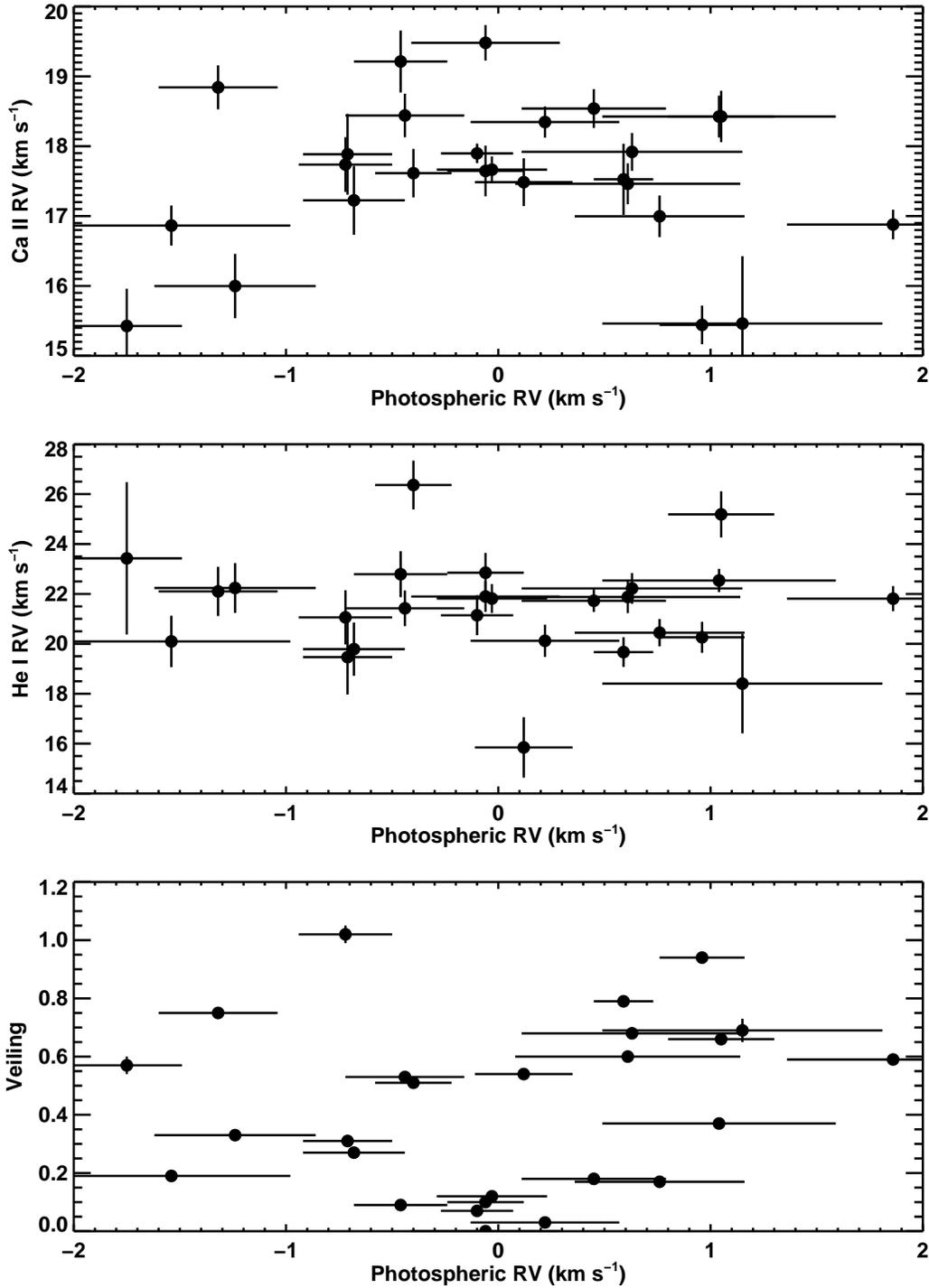}
\caption{The top panel shows the \ion{Ca}{2} 8662 \AA\ RV measurements 
versus the photospheric RV measurements for the optical data.  The middle
panel shows the \ion{He}{1} 5876 \AA\ RV values versus the photospheric
RVs, and the bottom panel shows the optical veiling versus the photospheric
RV values.  No significant correlation was found in any of these plots.}
\label{fig11}
\end{figure*}

\clearpage

\begin{figure*}
\includegraphics[angle=0,width=5.5in]{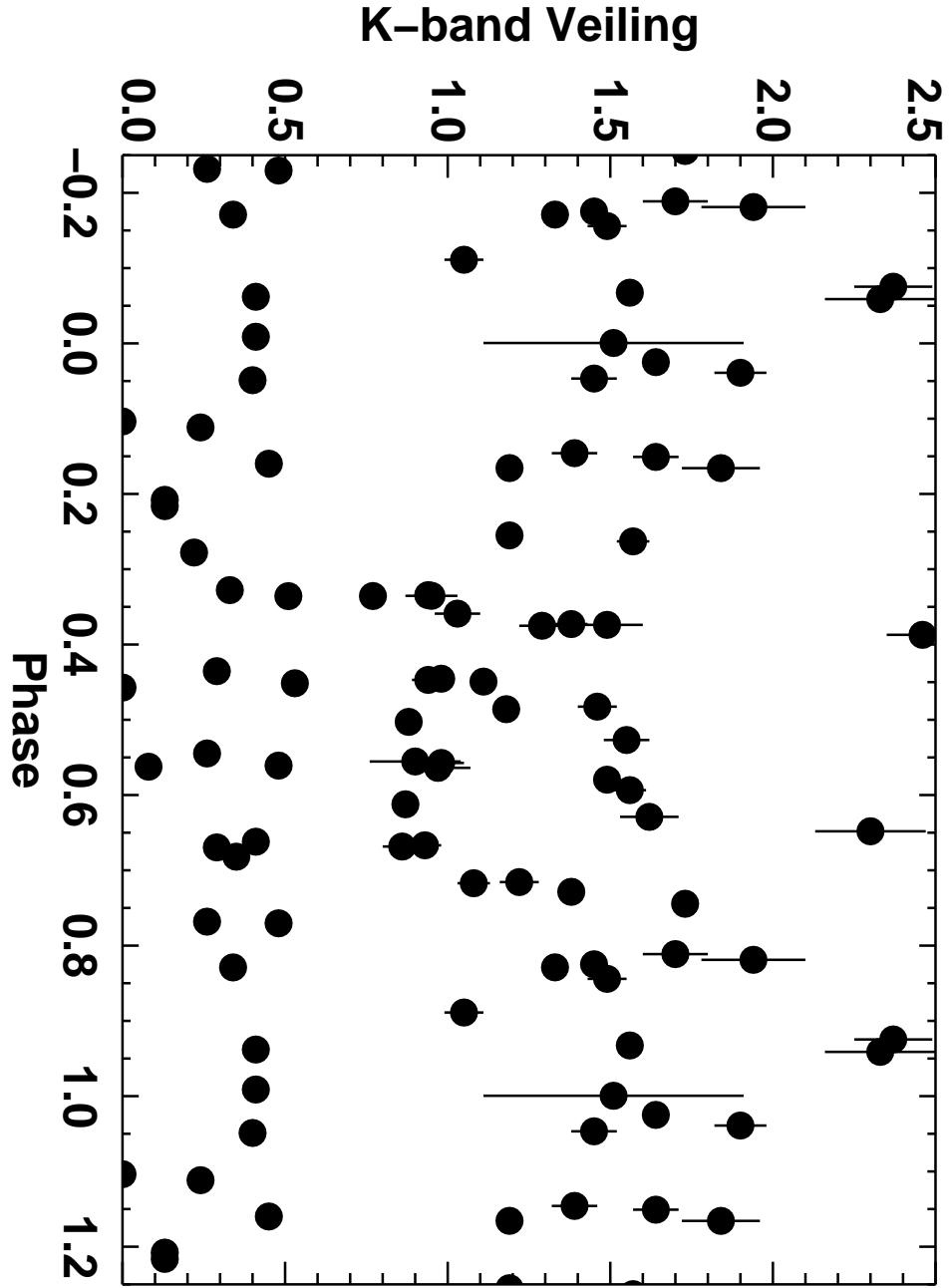}
\caption{The K-band veiling versus phase determined from the combined
optical plus IR RV fit (Figure 4).}
\label{fig12}
\end{figure*}

\clearpage

\begin{figure*}
\includegraphics[angle=0,width=5.5in]{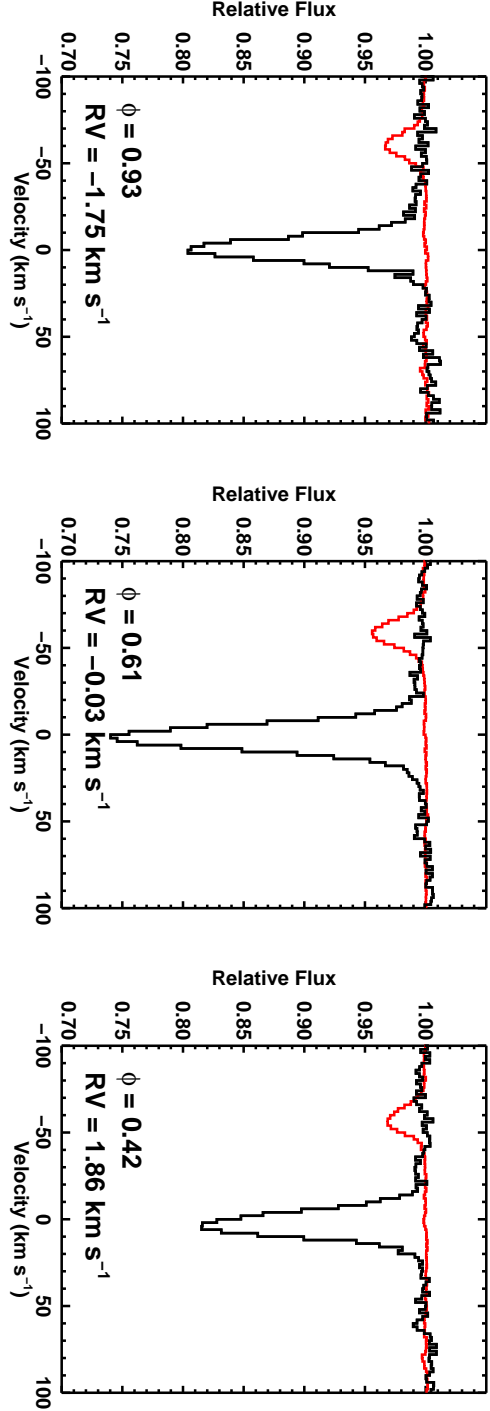}
\caption{Three representative LSD profiles from the optical spectra, showing
the observations with the smallest and largest measured RV as well as the
observation closest to zero RV.  The red profile shows the same profile 
offset by -60 km s$^{-1}$ and scaled to 17\% to show the strength of the
scattering feature needed to produce the measured RV variations (see text).
The phase ($\phi$) and RV of each observation is given.}
\label{fig13}
\end{figure*}

\end{document}